\documentclass{emulateapj}
\usepackage{lscape}

\def\blender{{\tt BLENDER}}
\def\kepler{\emph{Kepler}}
\def\kms{\ifmmode{\rm km\thinspace s^{-1}}\else km\thinspace s$^{-1}$\fi}

\submitted{To appear in The Astrophysical Journal, 1 January 2011}

\shortauthors{Torres et al.}
\shorttitle{\blender}

\begin{document}

\title{Modeling \kepler\ transit light curves as false positives:
Rejection of blend scenarios for Kepler-9, and validation of
Kepler-9\,d, a super-Earth-size planet in a multiple system}

\author{
Guillermo Torres\altaffilmark{1},
Fran\c{c}ois Fressin\altaffilmark{1},
Natalie M. Batalha\altaffilmark{2},
William J. Borucki\altaffilmark{3},
Timothy M. Brown\altaffilmark{4},
Stephen T. Bryson\altaffilmark{3},
Lars A. Buchhave\altaffilmark{5},
David Charbonneau\altaffilmark{1},
David R. Ciardi\altaffilmark{6},
Edward W. Dunham\altaffilmark{7},
Daniel C. Fabrycky\altaffilmark{1},
Eric B. Ford\altaffilmark{8},
Thomas N. Gautier III\altaffilmark{9},
Ronald L. Gilliland\altaffilmark{10},
Matthew J. Holman\altaffilmark{1},
Steve B. Howell\altaffilmark{11},
Howard Isaacson\altaffilmark{12},
Jon M. Jenkins\altaffilmark{13},
David G. Koch\altaffilmark{3},
David W. Latham\altaffilmark{1},
Jack J. Lissauer\altaffilmark{3},
Geoffrey W. Marcy\altaffilmark{14},
David G. Monet\altaffilmark{15},
Andrej Prsa\altaffilmark{16},
Samuel N. Quinn\altaffilmark{1},
Darin Ragozzine\altaffilmark{1},
Jason F. Rowe\altaffilmark{3,17},
Dimitar D. Sasselov\altaffilmark{1},
Jason H. Steffen\altaffilmark{18},
William F. Welsh\altaffilmark{19}
}

\altaffiltext{1}{Harvard-Smithsonian Center for Astrophysics, 
Cambridge, MA 02138, e-mail: gtorres@cfa.harvard.edu}
\altaffiltext{2}{San Jose State University, San Jose, CA 95192}
\altaffiltext{3}{NASA Ames Research Center, Moffett Field, CA 94035}
\altaffiltext{4}{Las Cumbres Observatory Global Telescope, Goleta, CA 93117}
\altaffiltext{5}{Niels Bohr Institute, Copenhagen University, DK-2100 Copenhagen, Denmark}
\altaffiltext{6}{NASA Exoplanet Science Institute/Caltech, Pasadena, CA USA 91125 }
\altaffiltext{7}{Lowell Observatory, Flagstaff, AZ 86001}
\altaffiltext{8}{University of Florida, Gainesville, FL 32611}
\altaffiltext{9}{Jet Propulsion Laboratory/California Institute of Technology, Pasadena, CA 91109}
\altaffiltext{10}{Space Telescope Science Institute, Baltimore, MD 21218}
\altaffiltext{11}{National Optical Astronomy Observatory, Tucson, AZ 85719}
\altaffiltext{12}{San Francisco State University, San Francisco, CA 94132}
\altaffiltext{13}{SETI Institute/NASA Ames Research Center, Moffett Field, CA 94035}
\altaffiltext{14}{University of California, Berkeley, CA 94720}
\altaffiltext{15}{US Naval Observatory, Flagstaff Station, Flagstaff, AZ 86001}
\altaffiltext{16}{Villanova University, Villanova, PA 19085}
\altaffiltext{17}{NASA Postdoctoral Program Fellow}
\altaffiltext{18}{Fermilab Center for Particle Astrophysics, P.O. Box 500, Batavia, IL 60510}
\altaffiltext{19}{San Diego State University, San Diego, CA 92182, USA}

\begin{abstract}
Light curves from the \kepler\ Mission contain valuable information on
the nature of the phenomena producing the transit-like signals. To
assist in exploring the possibility that they are due to an
astrophysical false positive, we describe a procedure (\blender) to
model the photometry in terms of a ``blend'' rather than a planet
orbiting a star.  A blend may consist of a background or foreground
eclipsing binary (or star-planet pair) whose eclipses are attenuated
by the light of the candidate and possibly other stars within the
photometric aperture. We apply \blender\ to the case of Kepler-9, a
target harboring two previously confirmed Saturn-size planets
(Kepler-9\,b and Kepler-9\,c) showing transit timing variations, and
an additional shallower signal with a 1.59-day period suggesting the
presence of a super-Earth-size planet.  Using \blender\ together with
constraints from other follow-up observations we are able to rule out
all blends for the two deeper signals, and provide independent
validation of their planetary nature. For the shallower signal we rule
out a large fraction of the false positives that might mimic the
transits. The false alarm rate for remaining blends depends in part
(and inversely) on the unknown frequency of small-size planets. Based
on several realistic estimates of this frequency we conclude with very
high confidence that this small signal is due to a super-Earth-size
planet (Kepler-9\,d) in a multiple system, rather than a false
positive. The radius is determined to be
$1.64^{+0.19}_{-0.14}\,R_{\earth}$, and current spectroscopic
observations are as yet insufficient to establish its mass.
\end{abstract}


\section{Introduction}
\label{sec:introduction}

The \kepler\ Mission, launched in March of 2009, was designed to
address the important question of the frequency of Earth-size planets
around Sun-like stars, and to characterize extrasolar transiting
planets through a 3.5-year program of very precise photometric
monitoring of $\sim$156\,000 stars \citep{Koch:10}. Science results
from the mission have already begun to appear \citep{Borucki:10a,
Borucki:10b, Steffen:10}. As shown already by ground-based surveys for
transiting planets, considerable effort is required to validate
candidates detected photometrically. This is because false positives
usually outnumber true planetary systems by a large factor, which is
about 10:1 for the most successful surveys from the ground, but is not
yet well characterized for \kepler.  The follow-up efforts by the
\kepler\ team have been summarized by \cite{Batalha:10}.

Spectroscopy is often a crucial step in the vetting process, as it
allows not only to measure the mass of a planet but also to examine
any changes in the line profiles (bisector spans) that might indicate
a false positive \citep[see][]{Queloz:01, Torres:05}. Some of the most
challenging false positives to rule out include chance alignments with
a background eclipsing binary (``blends'').  However, for faint
candidates ($V > 14$) high-resolution, high signal-to-noise ratio
spectroscopy becomes prohibitively expensive in terms of telescope
time.  Even for brighter candidates, the reflex motion of the star due
to an Earth-mass planet can sometimes be below the radial-velocity
detection limit, making spectroscopic confirmation very difficult or
impossible.  The question then becomes how to validate these
candidates, particularly the ones with small planets that are
precisely among the most interesting.

A number of other tests have been developed that can aid in
understanding the nature of the candidate, and that rely on the
long-term and nearly continuous photometric monitoring of \kepler, as
well as the very high astrometric precision achieved in determining
the centroids of the stars \citep[see also][]{Steffen:10}. These tests
include: \emph{i)} verifying that alternating events have the same
depth, which they may not if the signal is due to a background
eclipsing binary; \emph{ii)} checking for the presence of shallow
secondary eclipses, which are common in eclipsing binaries but are not
expected for the smallest planets; \emph{iii)} checking for
ellipsoidal variations, which could be another sign of a
blend. \emph{iv)} checking for changes in the centroid positions
correlated with the brightness changes, which, if detected, might
indicate a blend, or at the very least, a crowded aperture. This is a
powerful diagnostic that is able to disprove many background blends.

In addition to these tests, high-resolution imaging is an important
tool to identify neighboring stars that might be eclipsing binaries
with the potential to cause the transit-like signals.  The photometric
aperture of \kepler\ is large enough (typically many arc seconds
across) that it usually includes other stars in addition to the
candidate, which increases the risk of such blends.  In some cases,
near-infrared observations with Warm {\it Spitzer} can allow one to
reject the planet hypothesis if the transit depth at 3.6\,$\mu$m or
4.5\,$\mu$m is significantly different from that in the \kepler\
band. Such a signature might result from a blend with an eclipsing
binary of a different spectral type than the candidate.\footnote{For
Earth-size planets, the amplitude of the signal in the \kepler\ band is
very small and possibly below the detection threshold for {\it
Spitzer}.  However, a blend with a late-type binary could produce a
much deeper eclipse at longer wavelengths that may be detectable in
the near-infrared {\it Spitzer} bands.}

Even with this extensive battery of tests it may still be difficult or
impossible to provide validation for many of the most interesting
planet candidates discovered by \kepler. For example, blend scenarios
involving an eclipsing binary or an eclipsing star-planet pair
physically associated with the candidate (hierarchical triple systems)
and in a long-period orbit around their common center of mass would
often be spatially unresolved from the ground. These configurations
may also not be detectable spectroscopically, and would likewise not
produce any measurable centroid motion. Therefore, it is imperative to
take advantage of all the information available in vetting candidates.

With this as our motivation, we describe here the use of the \kepler\
light curves themselves in a different way to help discriminate
between true planetary transits and a large variety of possible blend
scenarios, on a much more quantitative basis than simple
back-of-the-envelope calculations could provide. The technique tests
these scenarios by directly modeling the light curves as blends, and
has considerable predictive power that allows the expected properties
of the various configurations to be tested against other information
that may be available. Both hierarchical triples and background blends
can be explored. A restricted application of this type of modeling to
\kepler\ has already been made for the five multi-planet candidates
announced recently by \cite{Steffen:10}. For the present paper we have
chosen to illustrate the full potential of the method, which we refer
to as \blender, by applying it to the unique case of \kepler\ Object
of Interest 377 (KOI-377, henceforth Kepler-9). This is a
multi-planet system reported and described in detail by
\cite{Holman:10}, with \emph{three} low-amplitude periodic signals in
its light curve. We have selected this system for two main reasons. On
the one hand, it represents the first unambiguous detection of transit
timing variations (TTVs) in an extrasolar planet, with a pattern of
variation seen in two of its signals (Kepler-9\,b and Kepler-9\,c)
that constitutes irrefutable evidence that the objects producing them
are bona-fide planets.  This offers an ideal opportunity to test
\blender\ because their true nature is already known.  On the other
hand, the third signal (KOI-377.03)\footnote{The name of this
candidate follows the convention of the \kepler\ Mission in which
individual transiting planet candidates are designated with a
numerical tag, and validated planets are given a \kepler\ number and
letter designation as in Kepler-9\,b.} is small enough that it would
correspond to a super-Earth, but validation of its planetary origin is
not yet in hand. Should it be validated, Kepler-9 would become an
even more remarkable laboratory for the study of the architecture of
planetary systems involving small planets.  Thus, exploring the wide
range of possible blend configurations that might mimic this shallow
signal is of the greatest interest for determining its true nature.

\kepler\ is likely to find many other candidate transiting planets
similar to KOI-377.03, for which final validation by other means is
not currently feasible, either because the expected radial-velocity
signal is too small, or because Doppler measurements are otherwise
complicated due to the star being chromospherically active, rapidly
rotating, or too faint. With the application to Kepler-9 we show
that our light-curve modeling technique is a powerful tool for
exploring astrophysical false positive scenarios that is complementary
to other diagnostics, and should play an important role in the
discovery of Earth-size planets around other \kepler\ targets.

\section{Simulating false positives with \blender}
\label{sec:blender}

In general the detailed shape of a light curve displaying transit-like
events can be expected to contain useful constraints on possible blend
scenarios that might be responsible for those signals. With photometry
of the quality of that provided by \kepler, those constraints can be
quite strong, and may be used to exclude many blend configurations and
provide support for the planetary hypothesis. It is thus highly
desirable to take advantage of this information, particularly since it
relies only on observations already in hand.

The idea behind \blender\ is to compare the transit photometry of a
candidate against synthetic light curves produced by an eclipsing
binary that is included within the photometric aperture of \kepler,
and is contaminating the light of the candidate. The usually deep
eclipses of the binary are attenuated by the light of the candidate,
and reduced in depth so that they appear transit-like. In principle
there is an enormous range of possible binary configurations that
could mimic all of the features of true planetary transits, including
not only their depth, but also the total duration and the length of
the ingress and egress phases. Generally it is only with detailed
modeling that these can be ruled out.  Possible scenarios include not
only background eclipsing binaries, but also hierarchical triples,
i.e., an eclipsing binary physically associated with the candidate in
a wide orbit around their common center of mass.

The basic procedure for simulating light curves with \blender\ was
described in detail by \cite{Torres:04}, and further changes and
enhancements are discussed below. Briefly, the brightness variations
of an eclipsing binary are generated with the binary light-curve code
EBOP \citep{Popper:81}, based on the Nelson-Davis-Etzel model
\citep{Nelson:72, Etzel:81}, and then diluted by the light of the
candidate for comparison with the \kepler\ observations. Effects such
as limb darkening, gravity brightening, mutual reflection, and
oblateness of the binary components are included.  The objects
composing the binary are referred to as the `secondary' and
`tertiary', and the candidate is the `primary'. The properties of each
object needed to generate the light curves (brightness and size) are
taken from model isochrones by \cite{Marigo:08}, parametrized in terms
of their stellar mass.\footnote{This particular set of isochrones was
chosen because it reaches lower masses than other models (nominally
0.15\,$M_{\sun}$, which we have extrapolated slightly for this
application to 0.10\,$M_{\sun}$, near the brown dwarf limit), and
because a convenient web tool provided by the authors allows easy
interpolation in both age and metallicity ({\tt
http://stev.oapd.inaf.it/cgi-bin/cmd}). Additionally, isochrone
magnitudes are available in a variety of passbands including the
\kepler\ and {\it Spitzer} passbands, as well as Sloan and
2MASS.\label{foot:models}}
For the primary star the appropriate isochrone is selected by using as
constraints the effective temperature, surface gravity, and
metallicity determined spectroscopically. We assign also a mass and a
radius from this isochrone, although these characteristics are
irrelevant for generating the model light curves. We then read off the
intrinsic brightness of the star (absolute magnitude) in the \kepler\
passband, which is the only property needed by \blender. The
brightness of the primary is held fixed throughout all simulations.
The parameters of the binary components are allowed to vary freely
over wide ranges in order to provide the best match to the \kepler\
photometry in a chi-square sense, subject only to the condition that
the two stars lie on the same isochrone, as expected from coeval
formation. To read off their properties (absolute magnitude and size)
we use the mass as an intermediate variable.  The specific isochrone
adopted for the binary depends on the configuration: for hierarchical
triple scenarios we adopt the same age and chemical composition as the
primary, whereas for background binaries the isochrone can be
different. The \kepler\ light curve itself does not provide a useful
constraint on the age or metallicity of the binary in the background
case, so a typical choice is a model for solar metallicity and a
representative age for the field such as 3\,Gyr.  For background
binary scenarios the distance between the binary and the main star is
parametrized for convenience in terms of the difference in distance
modulus, $\Delta\delta$. The inferred distance between the primary
star and the observer will vary from blend to blend because we
constrain the combined brightness of all components of the blend to
match the measured apparent brightness of the target. \blender\ is
also able to account for differential extinction between the primary
and the binary, which can have a non-negligible effect in some cases
given the relatively low Galactic latitude of the \kepler\ field.

Early versions of \blender\ have been used occasionally in recent
years to examine transiting planet candidates from ground-based
surveys such as OGLE, TrES, and HATNet \citep[see, e.g.,][]{Torres:04,
Torres:05, Mandushev:05, O'Donovan:06, Bakos:07}, as well as from
CoRoT \citep{Fressin:10}. These studies have exploited the predictive
power of \blender\ to estimate further properties of the blend
scenarios, by testing them against complementary information such as
color indices, optical/near-infrared spectroscopy, or near-infrared
photometry from {\it Spitzer}.  For the application to \kepler,
several important modifications have been made to \blender, including
the following:
\emph{i)} the ability to generate light curves integrated over the
29.4-minute effective duration of an observation when using
long-cadence data. This changes the shape of the transits
significantly, given the high precision of the \kepler\ photometry and
the relatively short timescales of the events
\cite[see][]{Gilliland:10, Kipping:10};
\emph{ii)} de-trending of the original \kepler\ light curves with a
1-day running median to remove instrumental effects, and rejection of
outliers;
\emph{iii)} the use of model isochrones specific to the \kepler\
passband, kindly computed for us by L.\ Girardi. \blender\ can now
also use proper limb-darkening coefficients for the same band, as
opposed to an approximation to the \kepler\ passband such as the
Johnson $R$ filter, which is considerably narrower.  \kepler\
limb-darkening coefficients have been computed by \cite{Sing:10} and
also A.\ Prsa;\footnote{\tt
http://astro4.ast.villanova.edu/aprsa/?q=node/8 \label{foot:prsa}}\,
\emph{iv)} extension to any optical or near-infrared passband. In
particular, for any scenario explored with \blender, light curves can
be computed at other wavelengths such as the 3.6\,$\mu$m and
4.5\,$\mu$m passbands of the IRAC instrument on Warm {\it Spitzer}, in
order to further test the blend hypothesis. Additionally,
\blender\ can predict the overall color of a blend in any pair of
passbands, including the effects of differential reddening for
background or foreground scenarios. These colors may then be compared
with the measured colors of a target. Extinction at different
wavelengths is computed following the prescription by
\cite{Cardelli:89};
\emph{v)} the ability to have the tertiary be a (dark) planet instead
of a star, in which case the corresponding free parameter becomes the
radius of the planet rather than the tertiary mass. The mass of the
planet has little effect on the light curves in most cases, but can
nevertheless be set to any value;\footnote{We note that this option of
\blender\ implicitly allows to consider white dwarfs as tertiaries, as
they are also Earth-size and contribute little light. Their mass is
significantly larger than a planet's mass, however, which in close
orbits can lead to distortions in the primary star causing ellipsoidal
variation. Gravitational microlensing may also occur in systems
involving white dwarfs with long enough periods, and may well be
detectable in the \kepler\ photometry.}
\emph{vi)} the ability to include extra light from other stars that
may be present in the \kepler\ aperture, which further dilutes the
intrinsic signatures from the eclipsing binary;
\emph{vii)} the ability to model systems with eccentric orbits.
Eccentricity changes the orbital velocities during transit, and can
therefore affect the size (mass) of stars that allow satisfactory fits
to the light curve.

When exploring blend scenarios involving hierarchical triple systems,
the free parameters of the problem are the mass of the secondary, the
mass of the tertiary (or its radius, if a planet), and the inclination
angle. A fourth variable, the difference in distance modulus, is added
for background blends. These quantities are typically stepped over
wide ranges in a grid pattern to fully map the $\chi^2$ surface.  For
the application to Kepler-9 below, stellar masses are allowed to vary
along the isochrones between 0.1\,$M_{\sun}$ and 1.4\,$M_{\sun}$,
although at the larger values the observed duration of the transits is
already difficult to match unless the events are highly grazing, in
which case the shape would be very different. For planetary tertiaries
the radii are allowed to be as large as 1.8\,$R_{\rm Jup}$\,; values
higher than this have not been observed.

\section{Application to Kepler-9}
\label{sec:koi377}

Kepler-9 (KIC\,3323887, 2MASS\,19021775+3824032) is a relatively
faint star compared to typical ground-based transit hosts (\kepler\
magnitude $Kp = 13.8$), which was observed by the mission beginning in
the first quarter of operations, and presents three distinct periodic
signals in its light curve. The two with the largest amplitudes have
periods of 19.24 days (Kepler-9\,b) and 38.91 days (Kepler-9\,c),
and brightness decrements of 6.5 and 6.0 mmag, respectively. The third
signal (KOI-377.03) is much shallower (0.2 mmag), and repeats every
1.59 days. The two longer periods are within 2.5\% of being in a 2:1
ratio, and both objects display obvious TTVs that are anti-correlated,
clearly indicating they are interacting gravitationally and therefore
orbit the same star, and are planetary in nature
\citep[see][]{Holman:10}.  The estimated radii are quite similar to
that of Saturn, and the masses are somewhat smaller than Saturn, based
on available radial-velocity measurements constrained by transit times
and durations.  The short-period signal has one of the smallest
amplitudes detected by \kepler, and may well correspond to a third,
super-Earth-size planet in the system, with an estimated radius of
only $\sim$1.5\,$R_{\earth}$ \citep{Holman:10}. However, because it
shows no TTVs related to the other two planets (nor is expected to, on
dynamical grounds), and is predicted to induce only a very small
reflex velocity on the parent star that may be below detection for
such a faint object, the true origin of this signal has not yet been
established.

In the absence of the crucial evidence of TTVs, each of the two
largest signals ---and indeed the third signal as well--- could in
principle be due to a different blend.\footnote{Unlikely as it may
seem to have three different blends operating in the same system, the
large photometric aperture, nearly uninterrupted monitoring, very high
photometric precision, and long-term coverage of \kepler\ coupled with
the large number of targets observed makes it more sensitive to
picking up odd cases such as this, so they should not be completely
ruled out. An example already exists among the five multi-planet
candidates recently reported by \cite{Steffen:10}, in which one of the
systems (KOI-191) presents three transit-like signals, and one of
those signals (0.4 mmag depth) has been shown to be due to a
background eclipsing binary 2.6 mag fainter than the target, located
1.5 arcsec away.}  Therefore, as an illustration of the application of
\blender, we model the light curve of Kepler-9 at each period
separately, as we would any candidate with a single period, and we
account for possible blends at the other periods by incorporating
extra dilution consistent with those other scenarios. The goal for the
two largest signals is to demonstrate, as a sanity check, that
\blender\ would be able to rule out blends in similar cases where
confirmation is lacking, which \kepler\ is expected to find in
significant numbers. For the third signal of unknown nature, the
application of \blender\ should provide valuable evidence one way or
the other.

\subsection{Stellar properties and photometry}
\label{sec:properties}

Kepler-9 is a solar type star. The spectroscopic properties of the
primary are adopted from \cite{Holman:10}: $T_{\rm eff} = 5777 \pm
61$\,K, $\log g = 4.49 \pm 0.09$, and [Fe/H] $= +0.12 \pm 0.04$. With
these parameters, a comparison with the stellar evolution models of
\cite{Marigo:08} yields a stellar mass of $M_{\star} = 1.07 \pm
0.05\,M_{\sun}$, a radius of $R_{\star} = 1.02 \pm 0.05\,R_{\sun}$,
and an age of about 1\,Gyr, along with the absolute magnitude in the
\kepler\ band.  Only the latter is used by \blender, and is held fixed
in our modeling. The distance to the star estimated from the same
models is about 650\,pc, ignoring extinction. Uncertainties in the
brightness of the primary stemming from errors in $T_{\rm eff}$, $\log
g$, and [Fe/H] are small. For example, the error in $\log g$, which
has the most direct influence on the intrinsic brightness, translates
to an uncertainty of little more than 0.1\,mag in the absolute
magnitude. This has an insignificant impact on our results.

The photometry used here consists of the long-cadence measurements
gathered for Kepler-9 during \kepler\ quarters~1, 2, and 3, spanning
218 days, and was treated slightly differently than indicated earlier
for a generic \kepler\ candidate because of the complications stemming
from the TTVs. Using the binary FITS tables from MAST (Multimission
Archive at STScI, {\tt http://archive.stsci.edu/kepler/}), the ``raw''
aperture photometry for each quarter was first de-trended using a
moving cubic polynomial fit robustly to out-of-transit data, with a
sliding window of 999 minutes before and after each individual
datapoint.  This technique removes long-term trends due to stellar
activity or instrumental errors, but retains the properties of each
transit light curve. Statistically significant outliers were removed.

For the two long-period signals, simple folding will not create an
accurate light curve because of the strong TTVs. Instead, we used a
``shift-and-stack'' technique, in which each transit event is
displaced so that it is centered at ``time'' zero using the measured
transit times from \cite{Holman:10}. Along with the measurements in
transit, nearly a full cycle of out-of-transit data were also shifted.
Specifically, we shifted nearly 25\% of an orbital cycle before the
transit, and nearly 75\% after the transit. This preserves any
curvature outside of eclipse, and in principle would also reproduce
any secondary eclipses, both of which can provide useful constraints
when modeling the light curve with \blender. We note, however, that
the strong TTVs of the transits would be accompanied by shifted
secondary eclipses in a way that can only be predicted by full
numerical integration. This shift-and-stack technique would not align
secondary eclipses correctly and thus their depth would need to be
significant in each individual event to be noticed.  There is no sign
of secondary eclipses at these periods in the data at the $10^{-4}$
level, as expected from the planetary nature of the objects, and thus
the failure of the shift-and-stack technique to correctly add up the
secondary eclipses does not affect our results.
After shifting, all the transit and out-of-transit data were
``stacked'' together and each data point was given a time relative to
time zero at the center of each transit event. This was done
separately for the 19-day and 39-day signals. We have been careful not
to use a full cycle of out-of-transit data to avoid using any
photometric measurements more than once in the input light curve.  

For the 1.6-day signal that repeats at regular intervals (since it
shows no TTVs), we created a light curve by simply masking out the
transits at the other two periods.

\subsection{Additional observations for false positive rejection}
\label{sec:observations}

\begin{figure}[b!]
\vskip 0.1in
\plotone{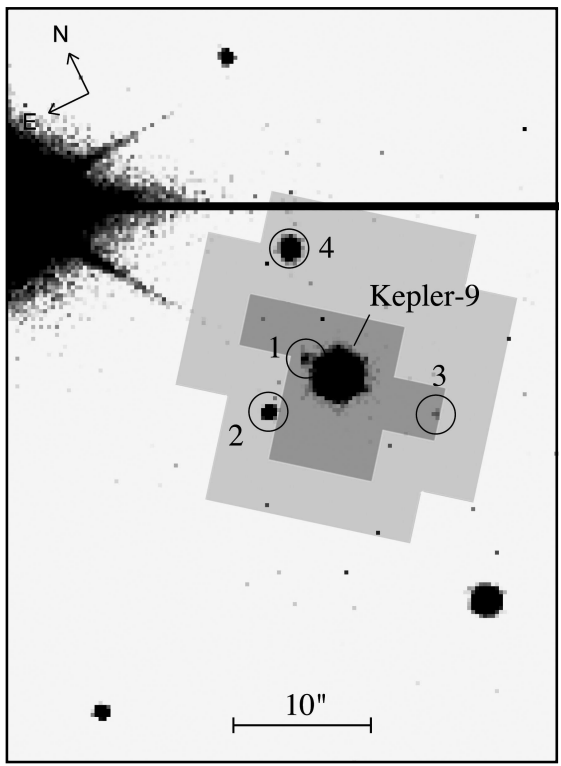}
\figcaption[]{Image of Kepler-9 from the HIRES guider camera on the
  Keck I telescope, obtained in seeing of 0\farcs9 and clear
  skies. Companions within 15\arcsec\ are labeled as in
  Table~\ref{tab:companions}. The scale of the image is 0\farcs30
  pix$^{-1}$. Also indicated are the optimal photometric aperture
  (darker gray area of 8 pixels, used to extract the \kepler\
  photometry) and the target aperture mask (lighter gray area of 31
  pixels, used to measure centroids) for \kepler\
  quarter~3.\label{fig:guider}}
\end{figure}

The photometric aperture of \kepler\ is typically a few pixels across,
with a scale of 3\farcs98 per pixel (see below).  High-resolution
imaging of Kepler-9 was performed in order to identify neighboring
stars that might be eclipsing binaries blended with and contaminating
the target photometry.  Images were recorded with the guider camera of
the High Resolution Echelle Spectrometer \citep[HIRES;][]{Vogt:94} on
the Keck~I telescope on Mauna Kea, in unfiltered light. The nominal
sensitivity of the CCD from 400 to 800 nm yields an effective passband
similar to the \kepler\ passband. The field of view was
43\arcsec$\times$57\arcsec, and the pixel scale was 0\farcs30 per
pixel.  One of these frames appears in Figure~\ref{fig:guider}, and
shows at least four stars in the field of view within 15\arcsec\ of
the target. Some of these stars are listed in various astrometric and
photometric catalogs. The brightness of these companions relative to
the target was measured using aperture photometry on four separate
Keck images, and ranges from $\Delta m = 2.6$ to 5.9 mag.

\begin{figure}
\epsscale{1.15}
\plotone{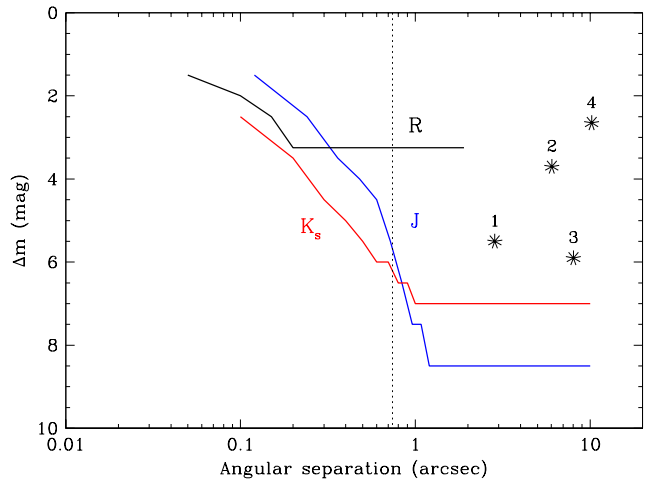}
\figcaption[]{Sensitivity to faint companions near Kepler-9 from our
imaging observations. Any companions above the curves are bright
enough to be detected. $J$ and $K_s$ limits are from AO observations
at the Palomar 200-inch telescope, and $R$ is from speckle
observations using the WIYN 3.5\,m telescope. Companions to the right
of the vertical dotted line at 0\farcs74 cannot be responsible for the
1.6-day signal, as they would have induced centroid motion that is not
observed. Stars detected in our imaging observations
(Table~\ref{tab:companions}) are marked with asterisks at their
measured angular separations and magnitude differences in the \kepler\
passband.\label{fig:AOsensitivity}}
%
\end{figure}

Speckle observations of Kepler-9 were carried out on 2010 June 18
with the WIYN 3.5\,m telescope located on Kitt Peak. They were taken
with a two-color EMCCD speckle camera using narrow-band filters 40\,nm
wide centered at 562\,nm and 692\,nm. We refer loosely to these
passbands as $V$ and $R$. The native seeing was 0\farcs7. No
companions with $\Delta m \leq 3.25$ mag ($R$ band) are present in the
field of view centered on the target out to 1\farcs8, at the 5-$\sigma$
confidence level. Inside of 0\farcs2 the sensitivity is reduced, but
still allows to rule out brighter companions down to the diffraction
limit of 0\farcs04--0\farcs05 (see Figure~\ref{fig:AOsensitivity}).
Details of the follow-up speckle observations in the context of the
\kepler\ Mission are described in more detail by \cite{Howell:10}.

\begin{deluxetable*}{lcccccc}[t!]
\tablewidth{0pc}
\tablecaption{Companions to Kepler-9 identified in our imaging observations.\label{tab:companions}}
\tablehead{
\colhead{Identification} &
\colhead{SDSS coordinates} & 
\colhead{$\rho$} & 
\colhead{P.A.} & 
\colhead{$\Delta J$} & 
\colhead{$\Delta K_s$} &
\colhead{$\Delta Kp$} \\
\colhead{} &
\colhead{(J2000)} &
\colhead{(\arcsec)} &
\colhead{(deg)} &
\colhead{(mag)} &
\colhead{(mag)} &
\colhead{(mag)}
}
\startdata
Kepler-9\tablenotemark{a} &  19:02:17.76 +38:24:03.2 & \nodata  & \nodata  & \nodata  & \nodata  & \nodata \\
Comp 1\tablenotemark{b}  &  19:02:17.91 +38:24:05.4 &   2.85   &   37.9   &   6.84   &   6.84   &   5.5   \\
Comp 2  &  19:02:18.27 +38:24:02.8 &   6.04   &   91.7   &   4.52   &   4.17   &   3.7   \\
Comp 3  &  19:02:17.29 +38:23:57.1 &   8.03   &  221.8   &   6.25   &   6.04   &   5.9   \\
Comp 4\tablenotemark{c}  &  19:02:17.69 +38:24:13.4 &  10.21   &  355.6   &   3.59   &   3.01   &   2.6 \\ [-1.5ex]
\enddata
\tablenotetext{a}{Target is also known as 2MASS\,19021775+3824032 and KIC\,3323887.}
\tablenotetext{b}{This companion is not listed in the SDSS catalog;
the coordinates are inferred from its position relative to Kepler-9.}
\tablenotetext{c}{Also known as 2MASS\,19021769+3824132 and KIC\,3323885.}
\end{deluxetable*}

Additionally, Kepler-9 was observed on 2010 July 2 at the Palomar
Hale 200-inch telescope with the near-infrared adaptive optics (AO)
PHARO instrument \citep{Hayward:01}, a $1024\times 1024$ Rockwell
HAWAII HgCdTe array detector. Observations were made in the $J$
(1.25\,$\mu$m) and $K_s$ (2.145\,$\mu$m) bands. The field of view was
approximately $20\arcsec\times20\arcsec$, and the scale was 25.1\,mas
per pixel. The AO system guided on the primary target itself, and
produced Strehl ratios of 0.05 at $J$ and 0.3 at $K_s$. The central
cores of the resulting point spread functions had widths of FWHM $=
0\farcs12$ at $J$ and FWHM $=0\farcs10$ at $K_s$. The closer of the
companions seen earlier in the Keck images were easily detected, and
we list them all in Table~\ref{tab:companions} along with relative
positions (angular separations and position angles), relative
brightness estimates, and other identifications.  The sensitivity to
faint companions was studied by injecting artificial stars into the
image at various separations and with a range of $\Delta m$, and then
attempting to detect them both by eye and with an automated IDL
procedure based on DAOPHOT. For firm detection we required the
artificial stars to be present in both passbands. The sensitivity
curves as a function of angular separation are shown in
Figure~\ref{fig:AOsensitivity}, along with the $R$-band sensitivity
estimated from the speckle observations.

Much fainter stars with $\Delta m > 9$ near a \kepler\ target could in
principle be detected by examining images from the Palomar Observatory
Sky Survey, which date back more than 50 years, provided the proper
motion of the target is large enough to have shifted it by several arc
seconds over that period. This is not possible for Kepler-9, since
its total proper motion as reported in the UCAC2 Catalog
\citep{Zacharias:04} is only 13.7 mas~yr$^{-1}$. The likelihood of
such faint close-in companions must therefore be addressed
statistically, if need be.

While the AO and speckle observations rule out the presence of bright
neighboring stars as close as 0\farcs1 or slightly less, further
limits on even tighter companions can be placed by the spectroscopic
observations obtained with HIRES on the Keck~I telescope, described by
\cite{Holman:10}, since those stars would fall well within the
0\farcs86 slit of the spectrograph. We performed simulations in which
we added the spectrum of a faint star to the original Kepler-9
spectra, over a range of relative brightnesses, and attempted to
detect these artificial companions by examining the cross-correlation
function.  We estimate conservatively that any such stars with
relative fluxes larger than about 10--15\% ($\Delta m$ less than
2--2.5 mag) would have been seen, unless their spectral lines are
blended with those of the target. The sharp lines of Kepler-9, with a
measured rotational broadening of only $v \sin i = 1.9 \pm 0.5$~\kms,
make this rather unlikely.

\subsection{Centroid analysis}
\label{sec:centroids}

Thanks to the very high astrometric precision of \kepler, an analysis
of the motion of the photocenter of a target provides an effective way
of identifying false positives that are caused by background eclipsing
binaries falling within the aperture. The principles have been
explained by \cite{Batalha:10} \citep[see also][]{Jenkins:10,
Monet:10}.  The centroid measurements described below use data from
quarter~3 only.  In quarter~1 the Kepler-9 aperture was determined to
be too small to optimally capture its flux, and was subsequently
enlarged. In quarter~2 \kepler\ experienced undesirable pointing
drift, which was later resolved. These problems complicate the
centroid analysis for quarters~1 and 2, although the results are
broadly consistent with the more reliable ones from quarter~3
presented here.

We describe first the use of difference image analysis to demonstrate
that the transit sources for all three Kepler-9 planets and
candidates are restricted to being very near the target star.  A
difference image is formed by averaging several exposures near, but
outside of a transit and subtracting from this the average of all
available exposures near transit center.  This results in a typically
isolated signal, a positive intensity with the shape of the point
spread function (PSF) at the \emph{true spatial location} of the
transit source, and an amplitude equal to the photometric transit
depth times the direct image intensity for the target.  Adopting 40
independent transits of KOI-377.03 in quarter~3 (avoiding those
shortly after major disturbances such as a safing event, and avoiding
any that overlap with `b' and `c' transits), each formed with six
symmetrically placed exposures outside of transits (after a two
exposure gap) and three near transit minimum, results in a 14-$\sigma$
signal in the difference image.  The corresponding direct image is
formed as the average of both in- and out-of-transit sets such that
the direct and difference images are sums and differences of precisely
the same exposure sets.

\begin{figure*}[t!]
\epsscale{0.8}
\plotone{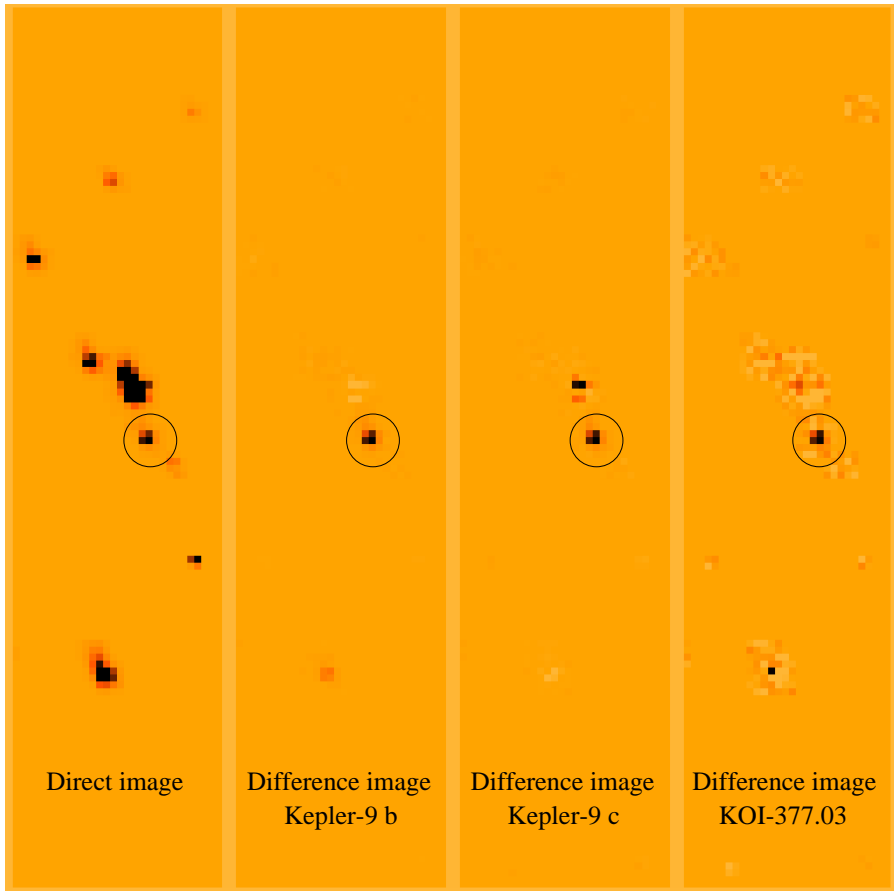}
\figcaption[]{Direct and difference images for Kepler-9.  The four
panels from left to right show 128 (row) by 30 (column) regions
corresponding to the direct and difference images for planets `b' and
`c', and candidate KOI-377.03.  The pixels returned for all stars in
this area have been mapped into original row and column locations on
the detector. Over 90\% of the image area is unfilled since \kepler\
returns only postage stamps on stars of interest.  The target
(KIC\,3323887) is indicated with circles in each panel.  The locally
brightest pixel is always at column 1100 and row 273, and each display
panel has been normalized by the sum of counts within the 3$\times$3
pixels centered on [1100, 273].  The display range is $-0.03$ to 0.3.
The difference images were created to isolate the signals for transits
`b', `c', and 377.03 respectively.  Most stars, not having variations
synced with these phases, effectively disappear in the difference
images.  For each of the three sets of transits the difference image
in the 3$\times$3 pixel core appears nearly identical to the direct
image, demonstrating that the true transit source must be near the
target to a small fraction of a pixel.  The difference images also
reflect the expected count levels for the source to be coincident with
the target.\label{fig:differenceimages}}
\end{figure*}

For Kepler-9\,b four transits were used from quarter~3 with five
exposures in transit, a gap of three, then five more exposures on each
side for out-of-transit.  Kepler-9\,c used two transits with seven
exposures in-transit, a gap of three, and seven symmetrically placed
out-of-transit exposure blocks.  By using only exposures pulled close
in time, and symmetrically with respect to the transits in use to form
a difference image, this effectively imposes a de-trending and avoids
any complications from drifts on time scales longer than the average
spread of the out-of-transit sets, which for Kepler-9\,c (the widest)
is about 9 hours.  Inspection of the difference images in
Figure~\ref{fig:differenceimages} shows that the transit sources for
the confirmed `b' and `c' planets \citep{Holman:10} and the candidate
KOI-377.03 must arise from close to the target star, with offsets
approaching one pixel easily ruled out by inspection.  A weighted PSF
fit \citep[or more properly, a Pixel Response Function (PRF) fit;
see][]{Bryson:10} to each of the direct and three difference images of
Figure~\ref{fig:differenceimages} is formed using only the central
3$\times$3 pixel area centered on the brightest pixel.  This leads to
offsets with respect to `b', `c', and KOI-377.03 of 0.007, 0.035, and
0.047 pixels, respectively.  For KOI-377.03 the formal error from a
weighted least squares fit is 0.062 pixels.  We have further assessed
the errors by generating a large number of independent realizations of
a transit signal of the KOI-377.03 relative intensity centered on the
target coordinates.  This leads to an {\em rms} scatter of 0.062
pixels.  The noise had been increased by a factor of 1.2 in the
difference image beyond direct Poisson plus readout noise estimates in
order to yield this congruence of least-squares errors and scatter in
simulations.  The distribution of offsets follows expected Gaussian
statistics, e.g., in the 7472 trials for KOI-377.03 the extreme offset
is 4.2$\sigma$ compared to the expected 4.0.  We have also shown that
simulating transit signals at 0.5 and 1.0 pixel offsets from the
target results in similar and smaller statistical scatter,
respectively, as less Poisson noise is under the transit image.  We
take the scatter of 0.062 pixels to generate a 3-$\sigma$ error circle
of 0.186 pixels, or 0\farcs74.  This is the minimum radius within
which background eclipsing binaries cannot be safely ruled out from
centroid analysis of the \kepler\ data itself.  To place this in
perspective: centroid analysis has ruled out 98.6\% of the area within
the 8-pixel optimal aperture ($>$99.6\% of the 31-pixel mask) of
Figure~\ref{fig:guider} as the location of potential background
eclipsing binaries creating the KOI-377.03 signal.

The quantitative results for all three transit sets are given in
Table~\ref{tab:differenceimages}.  Kepler-9\,b shows an offset of
0.0074 pixels between the difference and direct image relative to a
1-$\sigma$ error of 0.0049 pixels.  A 3-$\sigma$ error circle in which
background binaries cannot be excluded from the centroid analysis of
\kepler\ data itself is only 0\farcs06.  Kepler-9\,c is the only case
of the three showing a formal inconsistency with the offset being
5$\sigma$; however, even if we combine the offset and 3-$\sigma$
formal error any background eclipsing binaries outside of a radius of
0\farcs22 are excluded as the transit signal source.  Clearly for all
three transit sets, with the 3-$\sigma$ error circles comfortably
under 1\arcsec\ all of the known companions from high-resolution
imaging shown in Table~\ref{tab:companions} are safely ruled out as
sources of the photometric transit signal.  It is worth noting that
the formal (and equal to scatter from Monte-Carlo simulations) error
on radial offsets is approximately equal in pixel units to the inverse
photometric signal-to-noise (S/N) ratio, as expected \citep[see,
e.g.,][]{King:83}.

\begin{deluxetable}{lcccc}
\tabletypesize{\scriptsize}
\tablewidth{0pc}
\tablecaption{PRF centroid measurements on Kepler-9 direct and difference images.\label{tab:differenceimages}}
\tablehead{
\colhead{Type} &
\colhead{Intensity} & 
\colhead{Column} &
\colhead{Row} &
\colhead{Radius}
\\
\colhead{} &
\colhead{($e^-$)} & 
\colhead{(pix)} &
\colhead{(pix)} &
\colhead{offset (pix)}
}
\startdata 
\multicolumn{5}{c}{Kepler-9\,b} \\
\noalign{\vskip 2pt}
\hline
\noalign{\vskip 2pt}
 Direct       &  $5.170\times 10^7$ &     1099.6989     &     273.4557     & \\
 Difference   &  $4.360\times 10^5$ &     1099.7058     &     273.4584     & \\
 S/N ; Offset &    174              &   \phm{999}0.0069 &  \phm{99}0.0027  &   0.0074 \\
 Errors       &   2506              &   \phm{999}0.0029 &  \phm{99}0.0039  &   0.0049 \\
\noalign{\vskip 2pt}
\hline
\noalign{\vskip 4pt}
\multicolumn{5}{c}{Kepler-9\,c} \\
\noalign{\vskip 2pt}
\hline
\noalign{\vskip 2pt}
 Direct       &  $5.132\times 10^7$ &     1099.6999     &     273.4654     & \\
 Difference   &  $3.833\times 10^5$ &     1099.7107     &     273.4990     & \\
 S/N ; Offset &    128              &   \phm{999}0.0108 &  \phm{99}0.0336  &   0.0353 \\
 Errors       &   2991              &   \phm{999}0.0039 &  \phm{99}0.0056  &   0.0068 \\
\noalign{\vskip 2pt}
\hline
\noalign{\vskip 4pt}
\multicolumn{5}{c}{KOI-377.03} \\
\noalign{\vskip 2pt}
\hline
\noalign{\vskip 2pt}
 Direct       &  $5.401\times 10^7$ &     1099.6946     &     273.4408     & \\
 Difference   &  $1.493\times 10^4$ &     1099.7131     &     273.4836     & \\
 S/N ; Offset &    14.3             &   \phm{999}0.0185 &  \phm{99}0.0428  &   0.0047 \\
 Errors       &   1046              &   \phm{999}0.0351 &  \phm{99}0.0511  &   0.0062 \\ [-1.5ex]
\enddata

\tablecomments{The first two lines of each block present Intensity and
two coordinate position PRF fit results for the Direct and Difference
images, respectively.  The third line shows the photometric
signal-to-noise for the intensity in the Difference image, then the
offset in position of the preceding two lines, with the last entry
being the quadrature sum of the column and row offsets.  Errors refer
to the PRF fit to the Difference image. The scale is 1 pixel =
3\farcs98.}

\end{deluxetable}

Further confirmation that at least the two deeper signals seen in
Kepler-9 are not due to known stars in the scene can be obtained by
placing simulated eclipses on the known stars in the aperture, and
comparing them with the observations.  The scene in the aperture is
modeled using stars in the \kepler\ Input Catalog
\citep[KIC;][]{Latham:05}, supplemented by the stars in
Table~\ref{tab:companions}.  All stars within a PRF size (15 pixels)
in row or column of Kepler-9's aperture are included.  To generate
the modeled out-of-transit image, the measured PRF is placed at each
star's location on the focal plane, scaled by that star's flux.  This
provides the contribution of each star to the flux in the aperture's
pixels.  For each star $s_i$ in the aperture, the depth $d_{s_i}$ of a
transit is computed that reproduces the observed depth in the aperture
pixels.  An in-transit image for each $s_i$ is created as in the
out-of-transit image, but with the flux of $s_i$ suppressed by $1-
d_{s_i}$.  These model images are subject to errors in the PRF
\citep[][]{Bryson:10}, so they will not exactly match the sky.

A flux-weighted centroid is computed for the out-of-transit image and
the in-transit-image generated for each star in the aperture.  This
produces row and column centroid offsets $\Delta R$ and $\Delta C$,
and the centroid offset distance $D = \sqrt{\Delta R^2 + \Delta C^2}$.

\begin{deluxetable*}{lccc}
\tabletypesize{\scriptsize}
\tablewidth{0pc}
\tablecaption{Observed centroid shifts for Kepler-9\,b, Kepler-9\,c, and KOI-377.03.\label{tab:measured_centroids}}
\tablehead{
\colhead{} &
\colhead{Kepler-9\,b} &
\colhead{Kepler-9\,c} &
\colhead{KOI-377.03}
}
\startdata
$\Delta R$ & \phs$2.52\times10^{-4} \pm7.78\times10^{-5}$ & \phs$1.65\times10^{-4} \pm9.34\times10^{-5}$ & $-1.24\times10^{-7} \pm6.19\times10^{-5}$  \\
$\Delta C$ & $-2.23\times10^{-4} \pm7.55\times10^{-5}$  & $-2.40\times10^{-4} \pm9.21\times10^{-5}$ & \phs$8.39\times10^{-6} \pm5.82\times10^{-5}$  \\
$D$ & \phs$3.41\times10^{-4} \pm7.66\times10^{-5}$  & \phs$2.91\times10^{-4} \pm9.25\times10^{-5}$ & \phs$8.39\times10^{-6} \pm5.82\times10^{-5}$  \\
$D/\sigma$ & $4.44$  & $3.15$ & $0.14$  \\ [-1.5ex]
\enddata

\tablecomments{The measurements are given in pixel units, and the scale is 3\farcs98 per pixel.}

\end{deluxetable*} 

\begin{deluxetable*}{lcc|lcc|lcc}[b!]
\tabletypesize{\scriptsize}
\tablewidth{0pc}
\tablecaption{Modeled centroid shifts due to transits on the known stars in the aperture with depths that
reproduce the observed depth.\label{tab:modeled_offsets}}
\tablehead{
\colhead{Object} &
\colhead{Modeled $D$} &
\colhead{$D/\sigma$} &
\colhead{Object} &
\colhead{Modeled $D$} &
\colhead{$D/\sigma$} &
\colhead{Object} &
\colhead{Modeled $D$} &
\colhead{$D/\sigma$}
}
\startdata
Kepler-9\,b & $1.22\times10^{-3} $ & 16.0 & Kepler-9\,c & $1.10\times10^{-3} $ & 11.9 & KOI-377.03 & $5.02\times10^{-5} $ &  0.86 \\
Comp 1 & depth $> 1$ &  \nodata & Comp 1 & depth $> 1$ &  \nodata & Comp 1 & $1.45\times10^{-4} $ &  2.49  \\
Comp 2 & $9.77\times10^{-3} $ &  127 & Comp 2 & $8.75\times10^{-3} $ &  94.6 & Comp 2 & $4.01\times10^{-4} $ &  6.89 \\
Comp 3 & depth $> 1$ &  \nodata & Comp 3 & depth $> 1$ &  \nodata & Comp 3 & $5.33\times10^{-4} $ &  9.16 \\
Comp 4 & $1.44\times10^{-2} $ &  188  & Comp 4 & $1.29\times10^{-2} $ &  139 & Comp 4 & $5.90\times10^{-4} $ &  10.1 \\ [-1.5ex]
\enddata

\tablecomments{Shifts are given in pixel units, and the scale is
3\farcs98 per pixel.  For Kepler-9\,b and Kepler-9\,c transits on
some companions can be ruled out because they require depth $>1$.}

\end{deluxetable*} 

To compare these modeling results with observation we must make
low-noise centroid measurements from the observed pixel data.  We do
this by creating out-of-transit and in-transit images from de-trended,
folded pixel time series.  For each pixel time series, the de-trending
operation has three steps: 1) removal of a median-filtered time series
with a window size equal to the larger of 48 cadences or three times
the transit duration; 2) removal of a robust low-order polynomial fit;
and 3) the application of a Savitzky-Golay filtered time series of
order 3 with a width of 10 cadences.  The Savitzky-Golay filter is not
applied within 2 cadences of a transit event, so the transits are
preserved.  The resulting pixel time series are folded with the
transit period.  Each pixel in the out-of-transit image is the average
of 30 points taken from the folded time series outside the transit, 15
points on either side of the transit event.  Each pixel in the
in-transit image is the average of as many points in the transit as
possible: seven for Kepler-9\,b and Kepler-9\,c, and four for
KOI-377.03.  Centroids are computed for the in-transit and
out-of-transit images in the same way as the modeled images.

Uncertainties of these centroids are estimated via Monte Carlo
simulation, where a noise realization is injected into 48-cadence
smoothed versions of the pixel time series for each trial. A total of
2000 trials are performed each for Kepler-9\,b, Kepler-9\,c, and
KOI-377.03.  The in- and out-of-transit images are formed using the
same de-trending, folding and averaging as the flight data.  The
measured uncertainties are in the range of a few times $10^{-5}$
pixels.

Table~\ref{tab:measured_centroids} shows the resulting measurements of
the centroids from quarter~3 pixel data, along with the
Monte-Carlo-based $1\sigma$ uncertainties.  The centroids are
converted into centroid offsets and offset distance with propagated
uncertainties.  Table~\ref{tab:modeled_offsets} shows the offset
distance $D$ predicted by the modeling method described above for each
target in the aperture.  We see that when the transit is on Kepler-9
itself we expect a measurable centroid shift for Kepler-9\,b and
Kepler-9\,c.  In this case the modeled centroid shift is about 3.7
times larger than that observed, though the signs of the offsets
agree.  This exaggeration of the centroid offset has been traced to
inaccuracies in the KIC used to create the model images.  Therefore,
the centroid shifts in Table~\ref{tab:modeled_offsets} should be
scaled by a factor 1/3.7.  If the transit were on one of the companion
stars in the aperture, then the modeled centroid shift would be an
order of magnitude larger than observed for Kepler-9\,b and
Kepler-9\,c, ruling out the companion stars as the source for these
signals.  Companion stars are not as definitively ruled out for the
KOI-377.03 transit by this technique.  After scaling the centroid
offsets as above, modeled transits on companions~3 and 4 have offsets
that are are about $2.5\sigma$, while companion~2 is $1.9\sigma$ and
companion~1 is less than $1\sigma$.  The modeled transit on Kepler-9,
however, is much smaller, consistent with the observed transit offset
for KOI-377.03.

\subsection{\blender\ analysis of Kepler-9 b and c}
\label{sec:koi377bc}

\begin{figure}[b!]
\epsscale{1.15}
\plotone{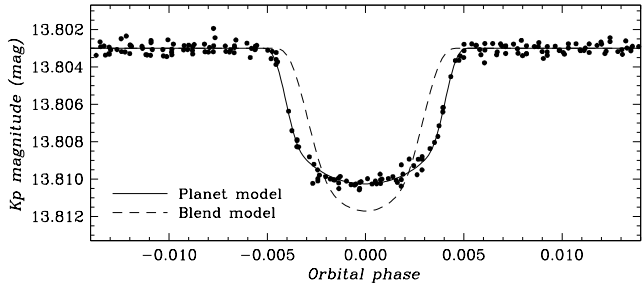}
\figcaption[]{Light curve of Kepler-9\,b ($P = 19.24$ days) with the
best fit blend model for the case of a hierarchical triple (candidate
+ physically associated eclipsing binary). The best fit planet model
is shown for reference. The poor fit of the blend model rules out this
configuration.\label{fig:koi377b_ht}}
\end{figure}

As an initial test, we modeled the light curves for each of these two
signals assuming they are the result of an eclipsing binary physically
associated with the target, i.e., at the same distance (hierarchical
triple). For this case the isochrone for the binary was taken to be
the same as that of the primary, and corresponds to [Fe/H] = +0.12 and
an age of 1~Gyr. The secondary and tertiary masses were allowed to
vary freely between 0.10\,$M_{\sun}$ (the lower limit in the models;
see footnote~\ref{foot:models}) and 1.40\,$M_{\sun}$, as mentioned
earlier, seeking the best fit to the photometry. The inclination angle
was also free, and the orbits were assumed to be circular. In both
Kepler-9\,b and c, which have similar transit signals, we find that
the best fitting hierarchical triple blend model corresponds to
secondaries that are approximately 1.0 and 0.5 mag fainter than the
primary, respectively, and tertiaries that are at the lower limit of
the isochrone range (late M dwarfs). However these fits give a poor
match to the photometry: \blender\ is unable to simultaneously
reproduce the total duration of the transit and the central depth,
given the constraints on the brightness and size of the stars from the
isochrones. This type of blend scenario is therefore clearly ruled
out. We illustrate this for Kepler-9\,b in
Figure~\ref{fig:koi377b_ht}, where the best-fit planet model is also
shown for reference. Much better matches to the data can be found if
additional light from a fourth star along the line of sight is
incorporated into the model, providing extra dilution. We find that
this fourth star is required to be nearly as bright as the primary,
and the optimal model changes in such a way that the secondary also
becomes as bright as the primary (so that its size enables the
duration of the transits to be reproduced), while the tertiary remains
a small star. This rather contrived scenario requiring two bright
stars that are nearly identical to the main star would be easily
recognized in our high-resolution imaging for separations larger than
about 0\farcs1 (see, e.g., Figure~\ref{fig:AOsensitivity}), in our
centroid analysis for separations larger than 0\farcs06, or would
otherwise produce obvious spectroscopic signatures unless all three
bright objects happened to have the same radial velocity.

\begin{figure}[t!]
\epsscale{1.25}
\plotone{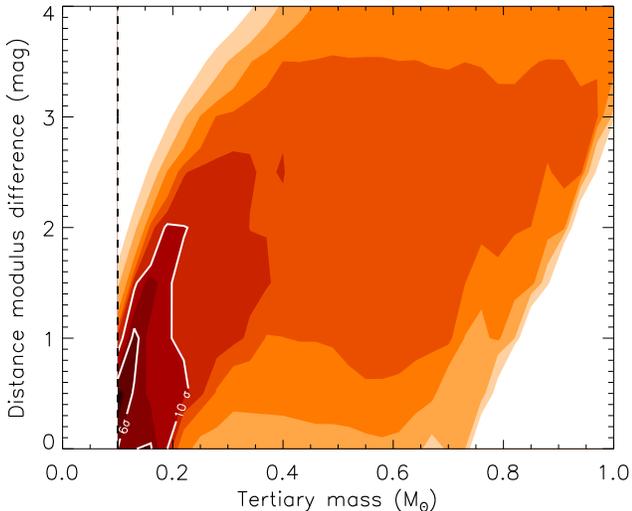}
\figcaption[]{Map of the $\chi^2$ surface (goodness of fit)
corresponding to a grid of blend models for Kepler-9\,c ($P = 38.91$
days) involving background eclipsing binaries with circular
orbits. The separation between the binary and the primary is expressed
in terms of the distance modulus difference. Contours are labeled with
the $\chi^2$ difference from the best planet model fit (expressed in
units of the significance level of the difference, $\sigma$), and are
plotted here as a function of the mass of the tertiary star. The
dashed line at 0.1\,$M_{\sun}$ indicates the lower limit to the
tertiary mass set by the model isochrones we
use. \label{fig:koi377c_backstar}}
\end{figure}

We next considered blends involving eclipsing binaries in the
background, by removing the constraint on the distance. In this case a
solar-metallicity isochrone was adopted for the binary, with a
representative age for the field of 3\,Gyr. We explored a wide range
of relative distances, and we first considered main-sequence stars
only, again with circular binary orbits.  The results for Kepler-9\,b
and c are once again similar to each other, and we illustrate them for
Kepler-9\,c in Figure~\ref{fig:koi377c_backstar}. The axes correspond
to the distance modulus difference $\Delta\delta$ as a function of the
tertiary mass. Contours represent constant differences in the $\chi^2$
of the fit compared to the best-fit planet model, and are labeled in
units of the statistical significance of the difference ($\sigma$). We
draw two main results from this figure. One is that the light curve
fits strongly prefer the smallest available tertiary masses from the
isochrones (0.10\,$M_{\sun}$), and would in fact yield better fits for
even smaller tertiaries (i.e., planets).  Additionally, the best
solutions cluster toward equal distances for the binary and the
primary star, effectively converging toward the equivalent of the
hierarchical triple scenario considered earlier. No acceptable
solutions exist with the binary at a significant distance behind the
primary star. The best fit to the light curve of Kepler-9\,c is
similar to the one shown in Figure~\ref{fig:koi377b_ht} (dashed
curve), which is not particularly good. The $\Delta\delta$ vs.\
tertiary mass diagram for Kepler-9\,b is qualitatively the same.
Allowing the secondary to be a giant star gives a very poor fit to the
photometry: the duration of the transit is very much longer than
observed, there is out-of-eclipse modulation due to distortions in the
giant, and all solutions place the binary at an implausibly large
distance.  We conclude that blend configurations involving background
eclipsing binaries in which the tertiary is a star are not a viable
explanation for either of these two signals.

\begin{figure}
\epsscale{1.25}
\plotone{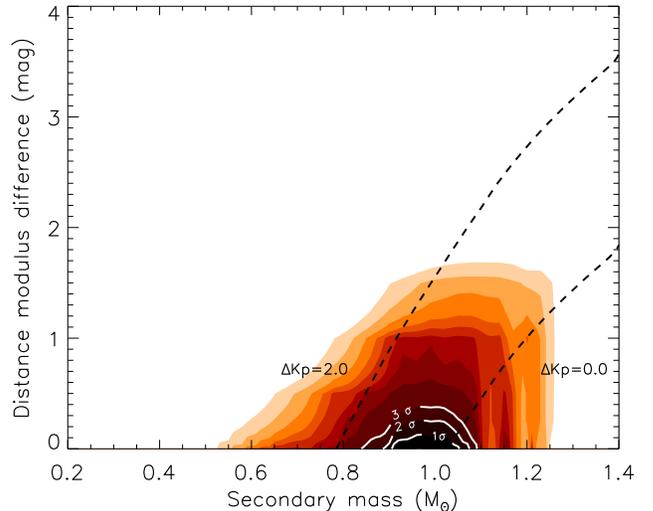}
%
\figcaption[]{Map of the $\chi^2$ surface corresponding to a grid of
blend models for Kepler-9\,c involving background eclipsing systems
in which the tertiary is a (dark) planet, in a circular orbit around
the secondary. Contours are labeled with the $\chi^2$ difference from
the best planet model fit (expressed in units of the significance
level of the difference, $\sigma$). Two dashed lines of equal
magnitude difference ($\Delta Kp$) are indicated, and show that all
viable blend fits (with confidence level $< 3\sigma$) have secondaries
that are bright enough to have been detected spectroscopically
($\Delta Kp < 2$).\label{fig:koi377c_backplanet}}
\end{figure}

We then explored background eclipsing binaries in which the tertiaries
are planets rather than stars. This allows their radii to be smaller,
possibly providing a better fit to the \kepler\ photometry. The orbits
were considered to be circular, as before.
Figure~\ref{fig:koi377c_backplanet} shows the results for
Kepler-9\,c, this time in the plane of separation versus secondary
mass. Once again the fits tends to favor an equal distance for the
binary and the primary star, and background scenarios with the binary
far behind provide unacceptably poor matches to the light curve. A
second noteworthy result is that these solutions have a strong
preference for secondary stars that are quite similar to the
primary. All acceptable fits to the light curve correspond to
relatively bright secondaries with $\Delta Kp < 1.5$ mag (see
Figure~\ref{fig:koi377c_backplanet}).  The best of these solutions is
of about the same quality as a planet model, and has a secondary of
mass 0.98\,$M_{\sun}$ that is only 100\,K cooler and 0.3 mag fainter
than the primary in the \kepler\ band.  This somewhat artificial case
of ``twin'' stars is a result we have seen often in simulations for
other \kepler\ candidates. The tertiary in this type of blend solution
comes out about $\sqrt{2}$ larger than in a planet model because the
transit is diluted by another star of approximately equal size and
brightness. One may debate whether this situation should actually be
referred to as a ``false positive'' for Kepler-9, since the signal
would still correspond to a gas giant planet, only that this planet
would be $\sqrt{2}$ larger, and it would be orbiting a different star.
Alternatively, it could be thought of simply as an overlooked dilution
factor in a true planetary system.  In any event, the lack of evidence
for this bright twin star in the spectroscopy or in our
high-resolution imaging or centroid analysis for Kepler-9 does not
support this scenario.

As a particular case of this family of configurations, we examined
blends in which the star-planet pair is constrained to be at the same
distance as the primary, i.e., effectively in a hierarchical system.
The secondary properties were therefore taken from the same isochrone
as the primary, and the orbits were assumed to be circular. An
excellent fit to the light curve is possible for a tertiary that is
about $\sqrt{2}$ larger than in a true planet model, but not
surprisingly, we find once more that the secondary must be as bright
as the primary.

Additional tests were run to examine the impact of changing the age
adopted for the isochrone of the secondary in a background star-planet
pair, or the addition of light from a fourth object in the
aperture. In the first case, changing the age from 3\,Gyr to 1\,Gyr
produced a small shift of the contours in
Figure~\ref{fig:koi377c_backplanet} downward and to the right that is
simply due to the change in intrinsic brightness of the secondary
star, and does not alter our conclusions.  Adding ``fourth light''
further attenuates the eclipses of the star-planet pair. To
compensate, \blender\ requires a slightly deeper eclipse, and in order
to preserve the shape of the signal (total duration, and slope of
ingress/egress), this is achieved by bringing the secondary closer to
the primary. As a result, for relatively bright fourth light the
contours are shifted downward by approximately the difference in
magnitude between the primary and the fourth star, again without
changing the conclusions.

\begin{figure}[b!]
\epsscale{1.15}
\plotone{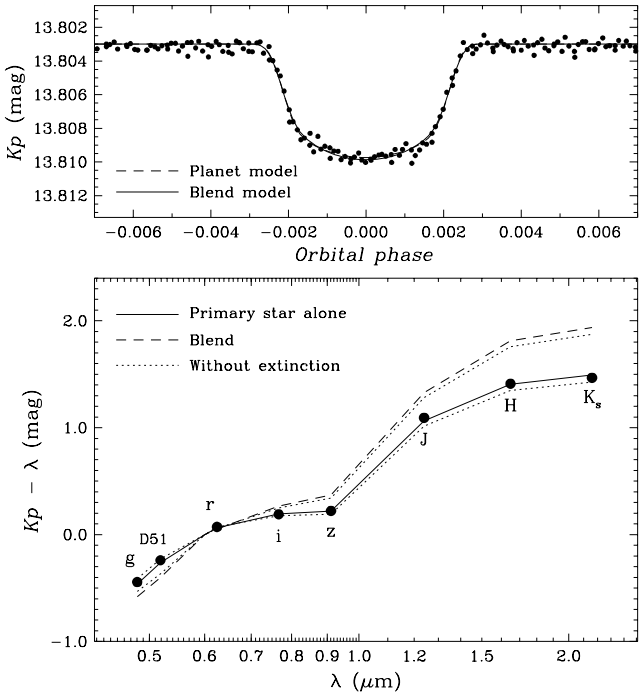}
\figcaption[]{\emph{Top}: Light curve of Kepler-9\,c with the best
fit blend model for the case of contamination by a foreground
eclipsing pair with a circular orbit in which the tertiary is a
planet. The pair consists of an M2 dwarf (0.56\,$M_{\sun}$,
0.58\,$R_{\sun}$) and a 0.91\,$R_{\rm Jup}$ companion 450\,pc in front
of the primary, which is at 750\,pc. This fit is statistically
indistinguishable from best fit planet model, also shown for
reference.  \emph{Bottom}: Measured colors for Kepler-9 (dots) compared
with the predictions from the blend model in the top panel. A small
amount of extinction (0.15 mag~kpc$^{-1}$) has been included in these
predictions. The results without considering extinction differ little,
and are shown with dotted lines. The color measurements clearly rule
out such a blend.\label{fig:colors}}
\end{figure}

One may also imagine blend scenarios in which the eclipsing binary is
in the foreground, rather than the background. We explored this
possibility by extending the simulations to negative values of
$\Delta\delta$. As before, we adopted circular orbits and a 3\,Gyr
isochrone for the foreground system. Binaries with stellar tertiaries
are clearly ruled out as they yield fits to the light curve that do
not match its shape, and additionally they predict a fairly obvious
secondary eclipse that is not seen in the data. We focus therefore on
blends in which the tertiary is a planet, and we illustrate the
results for Kepler-9\,c. In this case we find there are many
acceptable solutions with $\chi^2$ values differing from the best
planet fit at the level of 1$\sigma$ or less. These solutions span a
range of secondary masses and a range of foreground separations,
implying a wide range not only in apparent brightness for the
secondary, but also in color. Models in which the secondaries are
brighter than the primary and of significantly different spectral type
would be inconsistent with the spectroscopic parameters derived for
Kepler-9, and are excluded. Plausible solutions remain, in principle,
for fainter foreground secondaries, which necessarily involve
later-type stars. We find that of these, the only ones that yield
acceptably good fits to the \kepler\ photometry, with $\chi^2$ values
differing from the planet fit by less than 3$\sigma$, correspond to
secondaries that are within about 1.5 mag of the the primary in
brightness, and are of course redder. These would be valid blend
configurations so long as the secondaries are close enough to the
primary to be spatially unresolved (angular separations
$\lesssim0\farcs1$), and at the same time faint enough to have gone
undetected in the spectra. Stars that are within $\sim$2--2.5 mag of
the primary would generally have been seen spectroscopically, as
indicated in Sect.~\ref{sec:observations}, and this would exclude
these remaining foreground blend configurations.  Nevertheless, to be
conservative, let us assume for the moment that a star 1.5 mag fainter
than the primary has still managed to elude detection in our
spectra. This corresponds to the faintest secondary in a foreground
blend scenario that still allows for a satisfactory fit to the light
curve, and would be the most difficult case of this kind to
disprove. This fit is shown in the top panel of
Figure~\ref{fig:colors}, and is statistically indistinguishable from a
planet model fit. The secondary in this configuration is an M2 dwarf
($M = 0.56$\,$M_{\sun}$) 1.53 mag fainter than the primary, eclipsed
by a 0.91\,$R_{\rm Jup}$ planetary companion, and is located at a
distance of 300 pc. The primary in this scenario is at 750 pc.

Other properties of this particular blend such as magnitudes and
colors can be computed easily with \blender, and compared with
observations.  Apparent magnitudes for Kepler-9 are available from
the KIC for a variety of passbands including Sloan $griz$, a
special-purpose passband referred to as D51 (centered on the
\ion{Mg}{1}\,b triplet at 518.7\,nm), and $JHK_s$ from the 2MASS
catalog. The lower panel of Figure~\ref{fig:colors} shows various
color indices ($Kp-\lambda$) predicted by \blender\ both for the
primary star alone and for the blend. Those of the primary are well
reproduced by the model, and we find that a small amount of
interstellar extinction leads to an even better match (solid line in
the figure). The colors of the blend, on the other hand, disagree with
the measured colors, and deviate by more than 0.4 mag for the reddest
index, $Kp-K_s$.  We are therefore able to exclude, solely on the
basis of its color, this most difficult of the scenarios involving
foreground star-planet pairs that could mimic the 19-day and 39-day
signals in the light curve of Kepler-9.  Larger-mass secondaries
would not be as red and still allow for good fits to the photometry,
but they are intrinsically brighter and would be recognized more
easily.

\begin{figure}[b!]
\epsscale{1.15}
\plotone{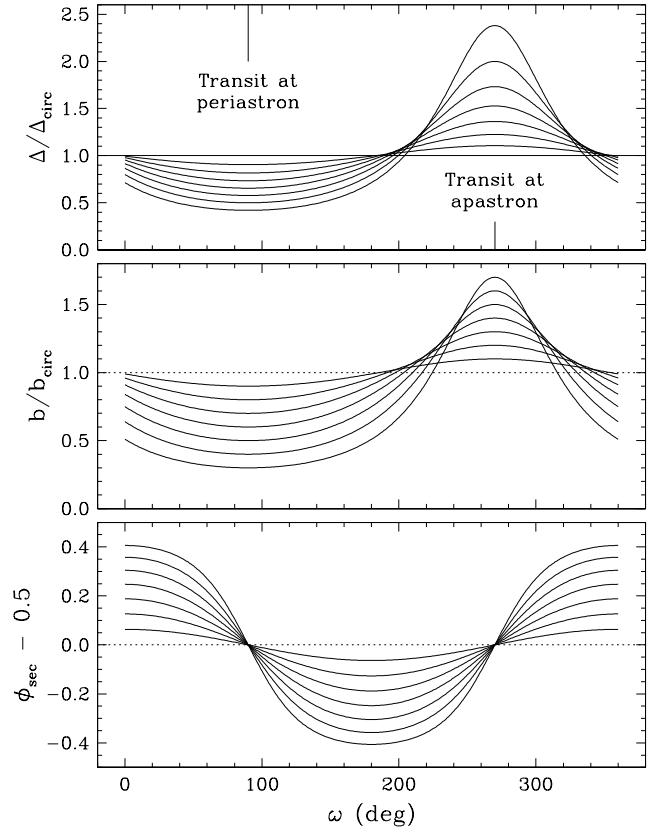}
\figcaption[]{Effect of eccentricity on the duration of transits
relative to the circular orbit case ($\Delta/\Delta_{\rm circ}$), on
the impact parameter ($b/b_{\rm circ}$), and on the displacement
($\phi_{\rm sec} - 0.5$) of the secondary eclipses relative to phase
0.5, all shown as a function of the longitude of periastron
$\omega$. The different curves correspond to eccentricities from 0.1
to 0.7 in steps of 0.1.\label{fig:ecc}}
\end{figure}

\begin{figure}[t!]
\vskip 0.1in
\epsscale{1.25}
\plotone{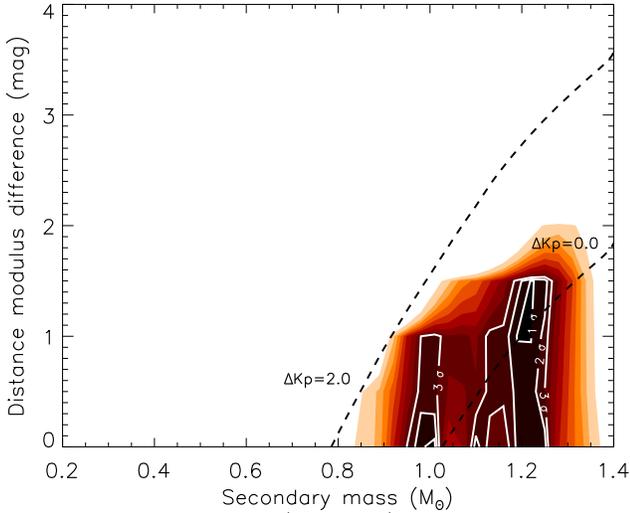}
%
\figcaption[]{Same as Figure~\ref{fig:koi377c_backplanet}
(Kepler-9\,c), restricted to star-planet orbits having $e = 0.3$
(concentration of contours on the left) and 0.5 (right), and $\omega =
90\arcdeg$. This orientation corresponds to transits that occur at
periastron.  Comparison with Figure~\ref{fig:koi377c_backplanet} shows
that these solutions allow for more massive (larger) secondary stars
than in the case of circular orbits, but the brightness of these
blends is still within 2 magnitudes of the target, and is ruled out by
spectroscopy.\label{fig:backplanet_ecc}}
\end{figure}

The above simulations have all assumed circular orbits for the blended
eclipsing binaries or star-planet pairs, which is not necessarily
realistic given the relatively long periods of Kepler-9\,b and c.
Eccentricity affects the speed of the secondary and tertiary in their
relative orbit, and therefore can change the duration of the transit,
making it shorter or longer than in a circular orbit, depending on the
orientation (longitude of periastron, $\omega$). It also changes the
impact parameter, all else being equal. And finally, it shifts the
location of the secondary eclipse. The magnitude of these effects is
illustrated in Figure~\ref{fig:ecc} for eccentricities between 0.1 and
0.7. The most important effect for our purposes is on the transit
duration.  Given a fixed (measured) duration, eccentric orbits may
allow blends with smaller or larger secondary stars than in the
circular case to still provide satisfactory fits to the light curve,
effectively increasing the pool of potential false positives. The
limiting cases correspond to $\omega = 90\arcdeg$ and 270\arcdeg, in
which the line of apsides is aligned with the line of sight and the
transit occurs at periastron (accommodating larger secondaries) or
apastron (smaller secondaries), respectively. Extensive simulations
for these two extreme situations show that allowing for eccentric
orbits does not change our conclusions regarding hierarchical triple
systems, background eclipsing binaries, or background star-planet
scenarios. We show this for the latter blend category in
Figure~\ref{fig:backplanet_ecc}, illustrated for the case of orbits
with eccentricities of 0.3 and 0.5, and $\omega = 90\arcdeg$.
Comparison with Figure~\ref{fig:koi377c_backplanet} indicates that in
both cases the blends are still bright enough that we would have seen
signatures of them in the spectra of Kepler-9. Larger eccentricities
of $e = 0.7$ result in secondaries that are brighter still. For
eccentric orbits oriented such that transits take place at apastron
($\omega = 270\arcdeg$), we only find acceptable fits to the light
curves for eclipsing star-planet pairs that are in the foreground (and
involve smaller stars). However, as was the case for circular orbits,
those blends are either too bright, too red, or both, and are thus
also excluded.

\begin{deluxetable*}{lcc}[b!]
\tabletypesize{\scriptsize}
\tablewidth{0pc}
\tablecaption{Summary of blend configurations tested for Kepler-9\,b and c.\label{tab:sumlarge}}
\tablehead{
\colhead{~~~~~~~~~~~~~~~~~~False positive configuration\tablenotemark{a}~~~~~~~~~~~~~~~~~~} &
\colhead{Result} & 
\colhead{Blends ruled out}
}
\startdata
Hierarchical triple with stellar tertiary, MS       &                           &      \\
~~~$\bullet$~Circular and eccentric orbits\dotfill  &   poor fits/sec.ecl.      & Yes  \\
~~~$\bullet$~Added fourth light\dotfill             &   twin star               & Yes (imaging/spec./centr.) \\ [+2.0ex]

Hierarchical triple with planetary tertiary, MS     &                           &      \\
~~~$\bullet$~Circular and eccentric orbits\dotfill  &   twin star               & Yes (imaging/spec./centr.) \\ [+2.0ex]

Background EB with stellar tertiary                 &                           &      \\
~~~$\bullet$~Circular and eccentric orbits, MS and giants\dotfill & poor fits   & Yes  \\ [+2.0ex]

Background EB with planetary tertiary, MS           &                           &      \\
~~~$\bullet$~Circular and eccentric orbits\dotfill  &   twin star               & Yes (imaging/spec./centr.) \\
~~~$\bullet$~1 Gyr isochrone for secondary\dotfill  &   little change           & Yes (imaging/spec./centr.) \\
~~~$\bullet$~Added fourth light\dotfill             &   little change           & Yes (imaging/spec./centr.) \\ [+2.0ex]

Foreground EB with stellar tertiary, MS             &                           &      \\
~~~$\bullet$~Circular and eccentric orbits\dotfill  &   poor fits/sec.ecl.      & Yes  \\ [+2.0ex]

Foreground EB with planetary tertiary, MS & & \\ 
~~~$\bullet$~Circular and eccentric orbits\dotfill & too bright/too red & Yes
(imaging/spec./centr./color) \\ [-1.5ex]
\enddata 

\tablenotetext{a}{3\,Gyr isochrone and solar metallicity assumed for
background and foreground stars, unless otherwise
indicated. Abbreviations: MS = main sequence secondary;
imaging/spec./centr. = high-resolution imaging, spectroscopy, and
centroid analysis; sec.ecl. = secondary eclipses predicted but not
observed; EB = eclipsing binary.}
\end{deluxetable*}

The above, fairly exhaustive exploration of parameter space with
\blender\ allows us to conclude that no configuration involving an
eclipsing binary (or an eclipsing star-planet pair), either in the
foreground or in the background, is able to provide a reasonable
explanation for the signals of Kepler-9\,b and c (see
Table~\ref{tab:sumlarge} for a summary of the configurations tested,
and the results). Many scenarios lead to light curves that match the
detailed shape of the transit events, but none are simultaneously
consistent with all of the other observational constraints. This
includes spectroscopy, high-resolution imaging, centroid measurements,
and photometry (colors). Therefore, even ignoring the evidence from
TTVs, these results fully support the planetary nature of these
objects and demonstrate the usefulness of \blender\ for validating
transiting planet candidates from \kepler.

\subsection{\blender\ analysis of KOI-377.03}
\label{sec:koi377.03}

We proceed next to examine false positive scenarios for the shallowest
signal in Kepler-9, with $P = 1.59$ days, which would correspond to a
super-Earth-size planet. Because the period is so short in this case,
and tidal forces in such binary systems have likely circularized the
orbit \citep[see, e.g.,][and references therein]{Mazeh:08}, we do not
consider non-zero eccentricities. Additionally, blends in which
the secondary star is a giant need not be considered, as those cases
are obviously ruled out because of the short orbital period and small
implied semimajor axis of the orbit.

\begin{figure}[b!]
\epsscale{1.25}
\plotone{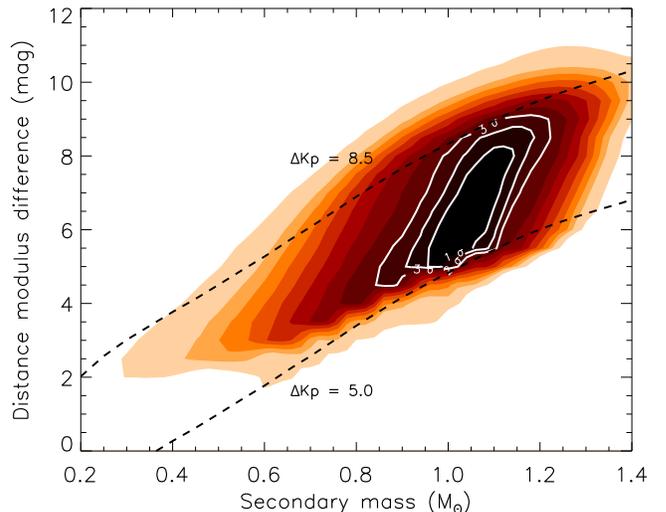}
%
\figcaption[]{Map of the $\chi^2$ surface for KOI-377.03 corresponding
to a grid of blend models involving background eclipsing binaries with
stellar tertiaries.  Contours are labeled with the $\chi^2$ difference
from the best planet model fit (expressed in units of the significance
level of the difference, $\sigma$). The dashed lines indicate levels
of equal apparent magnitude difference $\Delta Kp$ between the
background binary and the primary star.\label{fig:k3backstar}}
\end{figure}

As for the larger signals considered above, hierarchical triple
systems in which the tertiary is a star fail to provide good fits to
KOI-377.03.  A good match to the \kepler\ photometry can be found when
the tertiary is allowed to be a much smaller object (i.e., a planet),
but as was the case earlier, it requires a secondary that is very
similar to the primary in brightness. The resulting size of the
eclipsing object is $\sqrt{2}$ larger than in a planet model, or
slightly over 2\,$R_{\earth}$.  This type of configuration was ruled
out earlier based on the high-resolution imaging and the
spectroscopy. Small tertiaries with appreciable mass, such as white
dwarfs, induce tidal distortions on the primary due to the short
orbital period that lead to significant out-of-eclipse variations in
the light curve (ellipsoidal variability).  These modulations are not
seen in the photometry for Kepler-9, and such false positives are
therefore also excluded.

\begin{figure}[b!]
\epsscale{1.15}
\plotone{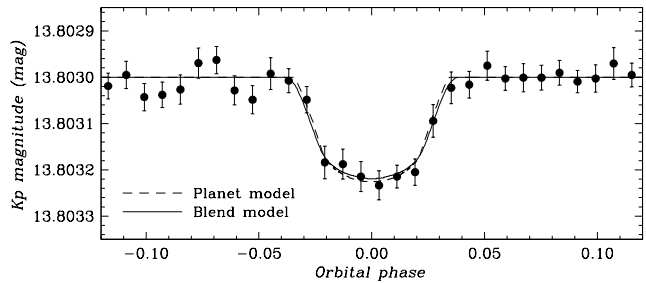}
%
\figcaption[]{Example of a blend model fit to KOI-377.03 involving a
background eclipsing binary with a stellar tertiary (solid line).  The
secondary is similar in spectral type to the primary and 5.2 mag
dimmer, and the tertiary is a late M dwarf. The eclipsing pair is
6\,kpc behind the primary. This fit is statistically indistinguishable
from the best fit planet model, which is shown with a dashed line.
The \kepler\ observations have been binned for
clarity.\label{fig:k3blend}}
\end{figure}

Blends with an eclipsing binary in the background (the tertiary being
a star) are able to reproduce the light curve just as well as a planet
model. In Figure~\ref{fig:k3backstar} we illustrate those results by
showing the area of allowed parameter space in a diagram of distance
modulus difference as a function of secondary mass. Acceptable fits
with $\chi^2$ differing from the planet model by less than 3$\sigma$
are possible over a wide range of relative separations ($4.5 \leq
\Delta\delta \leq 9$), but the secondaries are restricted to a
relatively narrow interval in mass centered on the mass of the
primary. These eclipsing binaries are all very distant and faint ($Kp
\approx 19$--22), and have no effect on the colors of the blend. The
more distant scenarios place the binary at implausibly large distances
of up to 42 kpc (more than 10 kpc above the Galactic plane). The
nearest configuration ($\Delta\delta = 4.5$; see
Figure~\ref{fig:k3backstar}) has the binary at a distance of 5.3\,kpc,
and the primary at $\sim$670 pc.  The secondary in this model is a
late G star 5.5\,mag fainter than the primary in the \kepler\
passband, eclipsed by a late M dwarf that produces no detectable
secondary eclipse. The predicted brightness of this binary precisely
matches that of the closest companion identified in the AO images
(Comp 1, Table~\ref{tab:companions}), located 2\farcs85 NE of the
target. However, this and all wider visual companions are already
ruled out at more than the 3-$\sigma$ confidence level by the lack of
centroid motion, which would have revealed any blended eclipsing
binaries at angular separations larger than about 0\farcs74
(Sect.\,\ref{sec:centroids}).  Even without this constraint from
astrometry, the predicted $J\!-\!K_s$ color of the secondary in this
blend is considerably redder than measured for this close AO
companion, which would also disqualify it.  Eclipsing binaries that
are between 5 and $\sim$8.5 magnitudes fainter than the main star
provide acceptably good fits to the light curve (see
Figure~\ref{fig:k3backstar}), and if they were angularly closer than
0\farcs74 from the target they may not be detectable in our AO or
speckle observations, in our centroid motion analysis, nor in our
spectra. They remain viable blend configurations, and would
necessarily be at distances greater than 5 or 6\,kpc. An example is
shown in Figure~\ref{fig:k3blend}, to illustrate that the fit is
indistinguishable from a planet fit.

\begin{figure}
\epsscale{1.25}
\plotone{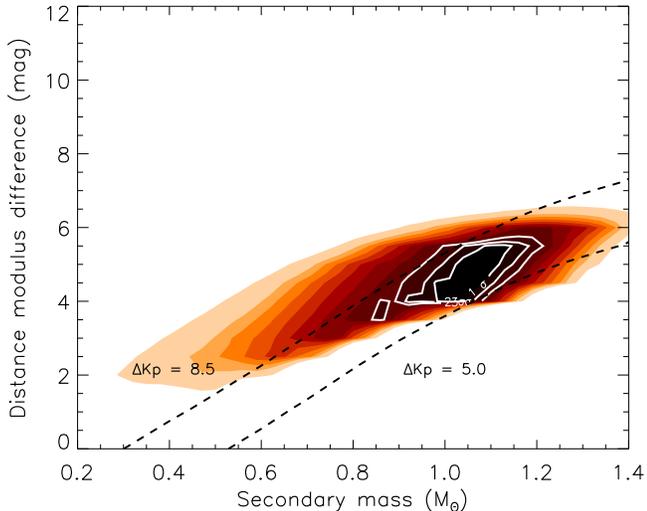}
%
\figcaption[]{Same as Figure~\ref{fig:k3backstar}, including the
effect of differential extinction in the amount of 0.5
mag~kpc$^{-1}$. The net effect of extinction is to compress and shift
the contours toward smaller relative distances. The dashed lines
indicate levels of equal apparent magnitude difference $\Delta Kp$
between the background binary and the primary star, and are the same
as shown in Figure~\ref{fig:k3backstar}.\label{fig:k3backstarextin}}
\end{figure}

In the above calculations we have ignored interstellar extinction.
However, given the large distances for the binaries in some of these
blend configurations, it is worth exploring the effect of dust more
carefully, which we have done by repeating the \blender\ simulations
using a representative differential extinction coefficient of 0.5
mag~kpc$^{-1}$. The results are shown in
Figure~\ref{fig:k3backstarextin}, and indicate that the blend
scenarios providing good fits to the \kepler\ photometry of KOI-377.03
are systematically shifted to smaller distances compared to the
previous calculations. Their apparent brightness, however, changes
relatively little, as can be seen by comparing the lines of equal
$\Delta Kp$ with those in Figure~\ref{fig:k3backstar}.  Therefore, the
overall impact of differential extinction on the permitted area of
parameter space in terms of observable parameters is not as
significant as might have appeared.

\begin{figure}
\epsscale{1.25}
\plotone{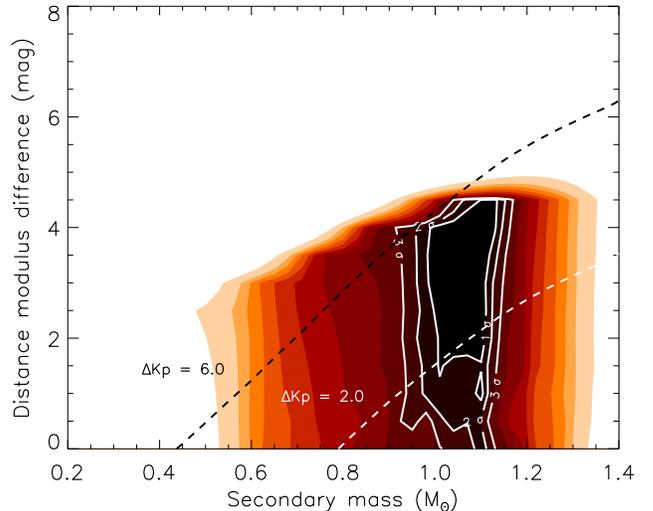}
%
\figcaption[]{Same as Figure~\ref{fig:k3backstarextin}, but for the
  case in which the tertiary is a planet instead of a
  star. Differential extinction is included. Kinks in the contours are
  an artifact of the discreteness of our grid. The dashed lines
  indicate levels of equal apparent magnitude difference $\Delta Kp$
  between the background secondary and the primary star. The lower of
  these lines represents the constraint from the spectroscopy for
  Kepler-9 (see text).\label{fig:k3backplan}}
\end{figure}

Allowing the tertiary to be a smaller object such as a planet opens up
a different area of parameter space for permissible background blends
(Figure~\ref{fig:k3backplan}). When including extinction as before,
acceptable fits to the light curve are possible for $\Delta\delta$
values from zero up to about 4.5.  This upper limit corresponds to
distances for the star-planet pair of about 4.8\,kpc (with $\Delta Kp
\approx 6$, or apparent magnitudes of $Kp \approx 20$), and is set by
the maximum size of 1.8\,$R_{\rm Jup}$ we have allowed for a
planet. As in the configurations described before, the solutions
constrain the secondary masses to be near that of the primary in order
to match the detailed shape and observed duration of the transits,
with a range from about 0.9\,$M_{\sun}$ to 1.2\,$M_{\sun}$. Therefore,
the color of the blend is not as useful a discriminant in this
case. Secondaries that are less than about 2 mag fainter than the
primary would have been seen spectroscopically. This excludes a good
fraction of the space of parameters, as indicated by the lower dashed
line in Figure~\ref{fig:k3backplan}. Of the remaining blends of this
kind between $\Delta Kp = 2$ and $\Delta Kp = 6$, only the ones with
angular separations smaller than about 1\arcsec\ are allowed by the
constraints from our AO imaging (see Figure~\ref{fig:AOsensitivity}),
but the centroid analysis is even more restrictive and rules out stars
outside of 0\farcs74.  At closer separations, the high-resolution
images rule out all blends that are brighter than the sensitivity
limit indicated in Figure~\ref{fig:AOsensitivity} (i.e., those that
fall above the curves).

\begin{figure}[t!]
\epsscale{1.25}
\plotone{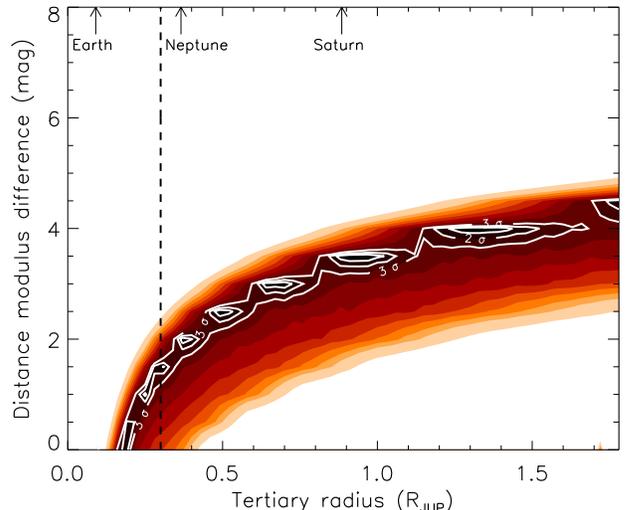}
%
\figcaption[]{Same as Figure~\ref{fig:k3backplan}, shown here as a
function of the tertiary radius. Kinks in the contours and closed
inner contours for the 1-$\sigma$ and 2-$\sigma$ levels are an artifact of the
discreteness of our grid. We indicate with a dashed line the lower
limit for the size of the tertiaries that is set by the spectroscopic
constraint on presence of bright stellar companions (see text). The size of
the Earth, Neptune, and Saturn are also indicated for
reference.\label{fig:k3backplanR3}}
\end{figure}

As expected from the fixed duration and depth of the transit-like
signal of KOI-377.03, the size of the tertiary in these configurations
correlates with the secondary mass.  Due to this correlation, small
tertiaries with $R \lesssim 0.3\,R_{\rm Jup}$ (roughly Neptune-size,
and smaller) are further excluded because the eclipses they produce
are already very shallow, and further dilution by the primary would
make them too shallow to fit the photometry. In order to avoid this,
the secondaries in those blends must be relatively small late-G type
stars that are nearby, and would therefore be bright enough ($\Delta
Kp \lesssim 2$) that they would have been detected in our spectra as a
second set of lines. Thus, \blender\ effectively places limits not
only on the secondary, but also on the size of the tertiary (see
Figure~\ref{fig:k3backplanR3}).  In particular, blends with a white
dwarf eclipsing a background star are also ruled out for the same
reason described above.  Additionally, the predicted light curves for
such cases with white dwarf tertiaries show ellipsoidal variability,
which is not observed.\footnote{We note, for completeness, that
the mass of a white dwarf would generally also be sufficient to induce
tidal synchronization in the secondary star, resulting in line
broadening that could in principle render it more difficult to detect
in the spectrum. However, given the orbital period and typical
secondary sizes, we estimate the rotational broadening to be no more
than $\sim$30\,\kms, which should still allow that star to be seen
spectroscopically if it were bright enough. In any case, white dwarfs
are excluded as viable tertiaries for the reasons mentioned in the
text.} Many of the larger tertiaries correspond to gas giants
(Saturn-size, or larger), which implies a qualitative difference in
their nature compared to the alternate model of a true Earth- or
super-Earth-size planet. In this sense these blends may properly be
considered ``false positives'', as opposed to the configurations
discussed earlier requiring twin stars, which only change the tertiary
radius by $\sqrt{2}$.

\begin{figure}
\epsscale{1.25}
\plotone{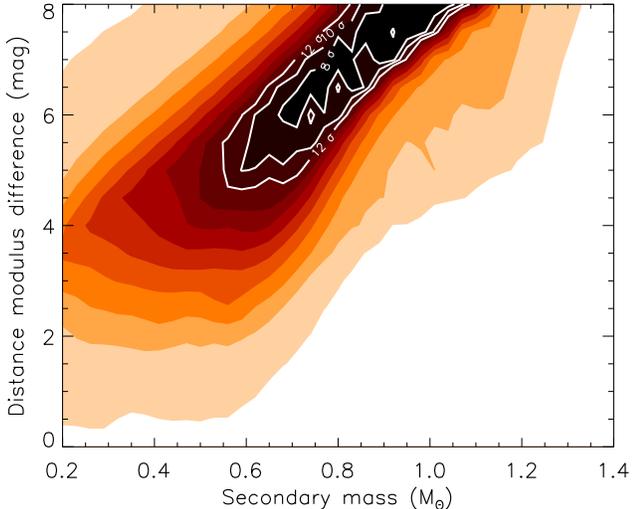}
%
\figcaption[]{Same as Figure~\ref{fig:k3backstar}, but for the case in
which the orbital period is assumed to be twice the nominal value ($2
P = 3.185702$ days).\label{fig:k3backstar2P}}
\end{figure}

\begin{figure}
\epsscale{1.15}
\plotone{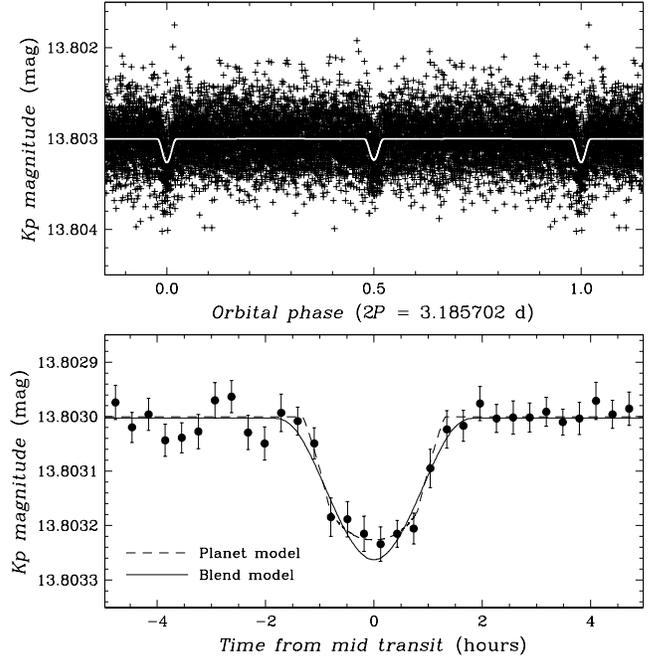}
%
\figcaption[]{Blend model of a background eclipsing binary with twice
the nominal period of KOI-377.03. \emph{Top}: \kepler\ observations
and best blend fit corresponding to two nearly equal mid G dwarfs
eclipsing each other, and located 20.4\,kpc behind the primary.
\emph{Bottom}: Binned observations compared against the blend model in
the top panel. The fit corresponding to a planet model is shown for
reference.\label{fig:2P}}
\end{figure}

Finally, we examine the possibility that the true period of the
KOI-377.03 signal is twice the nominal value. Alternating events would
then correspond to the primary and secondary eclipses of a blended
eclipsing binary (the tertiary being a star in this case), which may
in general be of different depth. In KOI-377.03 there is no compelling
evidence for a depth difference between odd- and even-numbered events,
but this is difficult to establish in a faint star such as this for a
signal that is only 0.2\,mmag deep. The results of extensive
simulations with \blender\ for this scenario are illustrated in
Figure~\ref{fig:k3backstar2P}.  The best fits correspond to blended
binaries far in the background, and do not indicate a significant
difference in depth between the primary and secondary eclipses.
However, these fits provide only a poor representation of the \kepler\
light curve, and can therefore be confidently ruled out. This is seen
in Figure~\ref{fig:2P}. The top panel shows the closest fit to the
full light curve together with the data, and in the bottom panel we
have binned the measurements to facilitate the comparison. This
solution involves an eclipsing pair of mid-G dwarfs at 21\,kpc and 7.2
mag fainter than the primary, and the fit is visibly worse than that
corresponding to a planet at half the period, which is shown with the
dashed line.

In summary, the \blender\ analysis of this section coupled with
constraints from spectroscopy, high-resolution imaging, and centroid
motion measurements rules out a large fraction of the false positives
that could produce the 1.59-day signal, but not all (see
Table~\ref{tab:sumsmall}). The remaining configurations involve
main-sequence background stars that are similar to the primary in
spectral type (late F to early K, or about 0.9\,$M_{\sun}$ to about
1.2\,$M_{\sun}$), and are eclipsed either by another main-sequence
smaller star or by a planet with $R_p > 0.3\,R_{\rm Jup}$
(Neptune-size or larger).  These blends range in apparent brightness
from $Kp \approx 19$ to $Kp \approx 22$ for stellar tertiaries, and
from $Kp \approx 16$ to $Kp \approx 20$ if the tertiaries are planets,
and must be closer than 0\farcs74 from the target. At separations
under 0\farcs74 our imaging observations allow us to rule out the
brighter of these blends, and only the ones with $\Delta m$ below the
sensitivity curves in Figure~\ref{fig:AOsensitivity} would remain
undetected.

\begin{deluxetable*}{lcc}
\tabletypesize{\scriptsize}
\tablewidth{0pc}
\tablecaption{Summary of blend configurations tested for KOI-377.03.\label{tab:sumsmall}}
\tablehead{
\colhead{~~~~~~~~~~~~~~~~~~~~False positive configuration\tablenotemark{a}~~~~~~~~~~~~~~~~~~~~} &
\colhead{Result} & 
\colhead{Blends ruled out}
}
\startdata
Hierarchical triple with stellar tertiary, MS\dotfill   &   poor fits               & Yes  \\
~~~$\bullet$~White dwarf tertiaries\dotfill             &   poor fits               & Yes  \\ [+2.0ex]

Hierarchical triple with planetary tertiary, MS\dotfill &   twin star               & Yes (imaging/spec./centr.)  \\ [+2.0ex]

Background EB with stellar tertiary, MS                 &                           &      \\
~~~$\bullet$~With and without extinction\dotfill        &   good fits               & Not all  \\
~~~$\bullet$~Giant secondaries\dotfill                  &   $P$ too short           & Yes  \\
~~~$\bullet$~White dwarf tertiaries\dotfill             &   poor fits               & Yes  \\
~~~$\bullet$~Twice the period\dotfill                   &   poor fits/sec.ecl.      & Yes  \\ [+2.0ex]

Background EB with planetary tertiary, MS               &                           &      \\
~~~$\bullet$~Jupiters, Neptunes, super-Earths, with extinction\dotfill & good fits & super-Earths \\
~~~$\bullet$~Giant secondaries\dotfill                  &   $P$ too short           & Yes  \\
~~~$\bullet$~White dwarfs, with extinction\dotfill      &   good fits               & Yes (imaging/spec./centr.) \\ [+2.0ex]

Foreground EB with stellar tertiary, MS\dotfill         &   poor fits               & Yes  \\ [+2.0ex]

Foreground EB with planetary tertiary, MS\dotfill       &   poor fits               & Yes  \\ [-1.5ex]
\enddata

\tablenotetext{a}{3\,Gyr isochrone and solar metallicity assumed for
background and foreground stars. Orbits are circular. Abbreviations:
MS = main sequence secondary; imaging/spec./centr. = high-resolution
imaging, spectroscopy, and centroid analysis; sec.ecl. = secondary
eclipses predicted but not observed; EB = eclipsing binary.}

\end{deluxetable*}

\subsection{Likelihood of remaining blend scenarios for KOI-377.03}
\label{sec:statistics}

In the previous section we have considered a wide variety of
possible blend scenarios for KOI-377.03 involving secondaries of
different spectral types (main-sequence stars and giants), eclipsing
objects of both planetary and stellar nature (including white dwarfs),
and configurations consisting of chance alignments with a foreground
or background contaminant, as well as hierarchical triple
systems. While these represent the most common and obvious
configurations one can imagine, in principle there could also be more
contrived scenarios that we have not thought of. These should be
intrinsically much less likely, a priori, but can nevertheless
not be completely ruled out. Therefore, we proceed below on the
assumption that any such situations we have not considered have a
small rate of occurrence, at least compared to the ones we have
discussed explicitly.

In order to provide the basis for an estimate of the confidence level
for the planetary status of KOI-377.03, we describe here the
calculation of the likelihood that the signal is due to a background
blend involving either a stellar tertiary or a planetary tertiary,
taking into account the constraints on brightness and other properties
indicated in the previous section. Because this type of calculation is
likely to be relevant for other \kepler\ candidates, we describe it
here in some detail.

The frequency of stars in the mass range permitted by \blender\ was
estimated using the Besan\c{c}on Galactic structure models of
\cite{Robin:03}, specifically for the $R$ band, which is the closest
available to the \kepler\ passband. We used an aperture of 1 square
degree centered on Kepler-9, and we performed the stellar density
calculations in half-magnitude bins of apparent brightness, accounting
for interstellar extinction as we did in the \blender\ simulations,
with a coefficient of 0.5~mag~kpc$^{-1}$ in $V$. The range of allowed
secondary masses for each magnitude bin was taken directly from
Figure~\ref{fig:k3backstarextin} for blends with stellar tertiaries,
and from Figure~\ref{fig:k3backplan} for blends with planetary
tertiaries.

Using the density of stars in each magnitude bin, we calculated the
fraction that would remain undetected after our high-resolution
imaging (speckle and AO observations), spectroscopy, and centroid
motion analyses. The results are listed in Table~\ref{tab:statistics}.
The first two columns give the $Kp$ magnitude range of each bin and
the magnitude difference $\Delta Kp$ compared to the target,
calculated at the upper edge of the magnitude bins.  For convenience
the calculations for blends with stellar tertiaries and planetary
tertiaries are listed separately. For the stellar tertiary case,
column~3 reports the density of stars obtained from the Besan\c{c}on
models, restricted to the mass range allowed by \blender\ as shown in
Figure~\ref{fig:k3backstarextin}.  Column~4 lists the maximum angular
separation $\rho_{\rm max}$ at which stars in the corresponding
magnitude bin would go undetected in our imaging observations, read
off from Figure~\ref{fig:AOsensitivity}, and taken at the center of
each magnitude bin. Centroid motion analysis rules out eclipsing
binaries beyond 0\farcs74, so $\rho_{\rm max}$ is constant at that
value for the last few bins in which this provides a stronger
constraint than the imaging limits.  The total number of stars of the
appropriate mass range in a circle of radius $\rho_{\rm max}$ around
Kepler-9 is then given for each bin in column~5, in units of
$10^6$. We note that the secondary mass and angular separation
constraints together reduce the number of background stars to be
considered as potential contaminants by a factor of $97,\!500$
compared to the number that would otherwise be expected to fall within
the photometric aperture of \kepler. This is already indicative of a
significantly reduced chance of having a false positive.

The intrinsic frequency of eclipsing binaries in the field is a key
ingredient in the calculation, and for this we have relied on the
results of \cite{Prsa:10}, which are based on the \kepler\
observations themselves.  These authors found the average occurrence
rate of eclipsing binaries among the \kepler\ targets down to $Kp
\approx 16$ to be approximately 1.2\% across the entire field. This
may be a slight overestimate because it counts as eclipsing binaries
targets that are actually blended with a background binary that is not
a target, though the effect is likely small.  There is little
information available on fainter eclipsing binaries, so we assume here
that a similar frequency holds. More importantly, many of these
eclipsing binaries cannot produce signals such as that of KOI-377.03
because their light curves have the wrong shape. Examples include
contact binaries and ellipsoidal variables, in which the brightness
changes continuously throughout the cycle, rather than presenting
sharp transit-like events such as we observe. Additionally,
semi-detached systems would have eclipses that are too long and also
of the wrong shape. We therefore exclude these from the tally.  With
this adjustment, the frequency of eclipsing binaries capable of
producing blends is 0.53\%. Multiplying the star counts in column~5 by
this frequency, we obtain the total number of blends expected in each
magnitude bin, which is reported in column~6.

Columns~7--9 are similar to columns~3--5, but for blends involving
star-planet pairs. Note that the range of allowed magnitudes is
different in this case, as is the range of secondary masses used to
compute the densities in column~7 (see Figure~\ref{fig:k3backplan}).
Following the size ranges adopted by \cite{Borucki:10b}, we consider
three categories of transiting planets as potential companions:
super-Earths (1.25--2\,$R_{\earth}$), Neptune-size planets
(2--6~$R_{\earth}$), and Jupiter-size planets (6--22 $R_{\earth}$, or
equivalently $\sim$0.5--2.0 $R_{\rm Jup}$).  Planetary tertiaries
smaller than 0.3\,$R_{\rm Jup} = 3.4\,R_{\earth}$ are ruled out by
\blender, as false positives with such tertiaries can only reproduce
the light curve if the secondaries are relatively bright, and those
would have been detected spectroscopically. This effectively
eliminates all super-Earths, and a fraction of the Neptunes. For the
intrinsic frequency of transiting planets we have relied on the
results from the first 43 days of \kepler\ observations as reported by
\cite{Borucki:10b}. That census is unlikely to have missed many
Neptune- or Jupiter-size planets (except ones with very long periods),
although it may include false positives, so we consider the count to
be conservative. Those authors presented a list of 306 targets with at
least one transiting planet candidate, and described another 400
targets with one or more transit-like signals that have not yet been
released. Of the 306 targets, 24\% would correspond to Jupiter-size
planets, and 23\% to Neptune-size planets. One may reasonably assume
that the 400 sequestered targets contain a larger fraction of smaller
Earth- or super-Earth-size planets (which cannot mimic the light curve
of KOI-377.03; see Sect.~\ref{sec:koi377.03}), for at least two
reasons. Smaller planets are the main focus of the \kepler\ Mission,
and they require more intensive follow-up efforts for validation,
which is why those targets have not yet been made public. Secondly,
these 400 targets are brighter, which makes smaller planets around
them easier to detect. Consequently, the assumption of similar
Jupiter- and Neptune-size planet frequencies as given above, but for
the entire sample of $306+400$ targets showing transit-like features,
is a conservative one. Scaling to the full sample of $156,\!097$
\kepler\ targets \citep{Borucki:10b}, we find an upper limit to the
frequency of transiting Jupiter-size and Neptune-size planets of
0.11\% and 0.10\%, respectively.\footnote{As a check we may compare
the above frequency of transiting Jupiter-size planets against results
from the statistical study by \cite{Fressin:09}. These authors
combined the findings of various radial-velocity searches and
folded-in the detections of very short-period systems detected by
ground-based transit surveys, which are typically less common in
Doppler searches. They found that 0.074\% of solar-type stars have a
transiting Jupiter-size planet. Our upper limit of 0.11\% is
consistent with this, given the presumably higher completeness of
\kepler\ and the fact that a fraction of the \kepler\ candidates that
is yet to be determined may turn out to be false positives once the
follow-up is completed.}  With these adopted frequencies, the
resulting numbers of blends involving these types of planets are
listed in columns~10 and 11 of Table~\ref{tab:statistics}.

The total blend frequencies in each of columns~6, 10, and 11 are
calculated as the sum of the individual frequencies in each magnitude
bin, and the three columns are then combined in the bottom section of
the table to yield an overall blend frequency (BF) of $\sim1.0 \times
10^{-7}$.

This very small figure corresponds to the number of false positives we
expect to find a priori for Kepler-9.  However, we point out
that this \emph{does not} translate directly into a false alarm
probability, or equivalently into a confidence level that the
candidate is orbited by a true super-Earth-size planet, as that
requires knowledge of the rate of occurrence of such planets.  For a
random candidate star in the field the rate of false positives
relative to the rate of true planets (false alarm rate, FAR) can be
written quite generally as ${\rm FAR} = N_{\rm FP}/(N_{\rm FP} +
N_p)$, where $N_{\rm FP}$ is the number of false positives and $N_p$
is the number of planets in the sample. Thus, the larger the number of
planets we expect, the smaller the FAR.

We consider Kepler-9 to be fairly representative of a typical target
in the field in terms of its spectral type (solar), brightness, and
background stellar density (a function of Galactic latitude). In that
case, the total number of blends can be taken to be approximately the
product of BF and the size of the sample, or $N_{\rm FP} = {\rm BF}
\times 156,\!097 = 0.016$.  The number of small planets expected in
the sample is of course not known, and determining it is precisely one
of the goals of the \kepler\ Mission.

If we accept a confidence level of 99.73\% (3$\sigma$) as being
sufficient for validation of a transiting planet candidate
(corresponding to ${\rm FAR} = 2.7 \times 10^{-3}$), then the minimum
number of super-Earth-size planets $N_p$ required in order to be able
to claim this level of confidence is six. Even though $N_p$ is
unknown, it is possible to make educated guesses as to what the
\emph{minimum} value would be in several ways, drawing on both
theoretical and observational considerations. 

Ground-based Doppler surveys continue to push toward the detection of
smaller and smaller planetary signals. \cite{Lovis:08} have reported
preliminary results from a sample of some 400 FGK stars observed with
the HARPS instrument on the ESO 3.6\,m telescope \citep[see
also][]{Mayor:09}, suggesting that as many as $\sim$30\% of the
targets may be orbited by close-in super-Earth- and Neptune-mass
companions (5--30\,$M_{\earth}$) with periods up to 50 days. The peak
in the period distribution seems to be around 10 days. By making use
of the CoRoTlux tool \citep{Fressin:07, Fressin:09} we simulated a
sample of $156,\!097$ stars in the \kepler\ field based on the
Besan\c{c}on models employed earlier, and used the results from
\cite{Lovis:08} to assign planets at random to each star, with a
log-normal distribution of periods centered at 10 days.  Approximate
planetary radii were inferred from the masses using the structure
models of \cite{Valencia:07}, by drawing masses at random and
assigning radii over the full range of compositions allowed by these
models.  We then retained only those in the super-Earth category, with
$R_p \leq 2\,R_{\earth}$.  The number of these planets that undergo
transits in the \kepler\ field is calculated to be $N_p \approx
200$. If we were to accept this estimate, the corresponding false
alarm rate would be ${\rm FAR} = 8 \times 10^{-5}$.

A similar Doppler survey of 166 G and K stars with the HIRES
instrument on the 10-m Keck~I telescope \citep{Howard:10} has provided
additional insights into the rate of occurrence of small-mass
planets. The results suggest that approximately 18\% of solar-type
stars harbor planets in the range of 3 to 30\,$M_{\earth}$ with
periods under 50 days. A calculation analogous to that carried out
above for the transit probabilities and conversion from planetary
masses to planetary radii leads to an estimate of $N_p \approx 120$
for the \kepler\ field. If we adopted this lower estimate, the
corresponding false alarm rate would be ${\rm FAR} = 1.3 \times
10^{-4}$.

Population synthesis studies such as those of \cite{Ida:04} and
\cite{Mordasini:09} based on the formation of planets by the core
accretion process and subsequent migration have provided tentative
predictions of the properties of planets, including distributions of
their masses, periods, and other characteristics. These theoretical
models seem to point to a sizable population of super-Earths that may
be several times larger than the number of Neptune- or Jupiter-mass
planets in those simulations. Scaled to the size of the \kepler\
sample, this would imply there could be several hundred small
transiting planets. However, the authors caution that those results
should be considered with great care as some of the physical
ingredients in these models are still very uncertain.

A further estimate may be obtained from the preliminary \kepler\
results as reported by \cite{Borucki:10b}. Among the 306 targets
listed there showing one or more periodic transit-like signals, 27
fall in the category of super-Earths. While it is true that these
candidates have not yet been followed up and validated, and may
therefore include some fraction of false positives, additional
super-Earth-size planets are to be expected in the list of 400
unreleased candidates, which could make $N_p$ considerably larger.
This is particularly true since the targets in the latter list are all
brighter than those in the publicly available set, and therefore the
proportion of small planets is likely to be higher because the transit
events are easier to detect. Nevertheless, if we were to accept that
$N_p$ is as small as 27, then the corresponding false alarm rate would
be ${\rm FAR} = 6 \times 10^{-4}$.

There are caveats associated with each of the observational estimates
mentioned above that should be kept in mind.  The \kepler\ results
invoked in the previous paragraph are still preliminary, and although
we regard our use of them to be conservative for the reasons described
earlier, the true fraction of false positives in the \kepler\ sample
remains unknown until all candidates have been followed up.  The
Doppler results are also preliminary to some degree, and are based on
somewhat limited samples of stars.  It is also possible that a
fraction of those Doppler candidates may turn out to be false
positives, or given the $\sin i$ ambiguity inherent in that technique,
that some of them may have actual masses above the range considered
for super-Earths.  Additionally, there are uncertainties associated
with the conversion we have applied between planetary masses and
planetary radii, using theoretical models. Those uncertainties are
difficult to quantify given our present state of knowledge.

For these reasons, added to the fact that despite our best efforts to
assess the blend frequency there could still be some exotic blend
scenario that we have overlooked, it is not possible to present a more
definitive value of the false alarm rate.  Nevertheless, the above
estimates of the FAR based on consideration of all the blend scenarios
that seem plausible to us are all sufficiently small that they give us
very high confidence that KOI-377.03 is not a false positive, and they
therefore validate it as a signal of planetary origin. We designate
this planet Kepler-9\,d.

\section{Discussion}
\label{sec:discussion}

Calculating the false alarm rate for targets with small signals such
as Kepler-9\,d is non-trivial because it depends crucially on the
frequency of small transiting planets, which may only be fully known
at the conclusion of the \kepler\ Mission.  Of the arguments for the
expected value of $N_p$ presented in the previous section, the one
that relies on the preliminary \kepler\ results themselves is the most
conservative, and already makes it highly unlikely that we are in the
presence of a false positive such as those explored in this paper.
Furthermore, that estimate is based on results from only the first
$\sim$43 days of operation of the spacecraft. Continued observations
over the next two years will surely increase the number of candidates,
which can only result in a larger confidence that the 1.6-day
photometric signal is due to a true planet. Thus, we find the
overall evidence for the planet interpretation very compelling. It is
also worth noting that the Doppler surveys have found that a very
large fraction of the smallest-mass planetary companions (as many as
80\%) are in multi-planet systems \citep[e.g.,][]{Lovis:08,
Mayor:09}. Because of the presence of Kepler-9\,b and c, this makes it
considerably more likely that Kepler-9\,d is also a planet than if it
were the only signal in the system.  Furthermore, one may expect
a priori that a planet interior to Kepler-9\,b and c would have a high
probability of presenting transits.  Indeed, with the reasonable
assumption that the orbit of the inner planet is more or less coplanar
with the outer two, the geometric probability of a transit at a period
of 1.6 days would be close to 100\%, instead of $\sim$18\% for random
inclinations.  

The light-curve parameters we obtain for Kepler-9\,d by modeling the
photometry using the formalism of \cite{Mandel:02} are summarized in
Table~\ref{tab:kepler9d}, and supersede the preliminary estimates of
\cite{Holman:10}.  The values and uncertainties were determined using
a Markov Chain Monte Carlo technique, with four chains of length
$10^6$ each. Our fits used the non-linear fourth-order limb-darkening
law of \cite{Claret:00}, with coefficients for the \kepler\ band taken
from the calculations by A.\ Prsa referenced in
footnote~\ref{foot:prsa}.  For an adopted stellar radius for the
primary star of $R_{\star} = 1.02 \pm 0.05\,R_{\sun}$, the estimated
size of this planet is $R_p = 1.64^{+0.19}_{-0.14}\,R_{\earth}$, which
is among the smallest yet reported.\footnote{The size of Kepler-9\,d
is not significantly different from that of CoRoT-7\,b, which is $R_p
= 1.68 \pm 0.09$\,$R_{\earth}$ according to \cite{Leger:09}, and was
revised to $R_p = 1.58 \pm 0.10\,R_{\earth}$ by \cite{Bruntt:10}. The
measured radius of the next smallest known planet, GJ\,1214\,b, is
$R_p = 2.68 \pm 0.13$\,$R_{\earth}$ \citep{Charbonneau:09}.}  The
uncertainly is currently dominated by the photometric errors, rather
than the stellar parameters, and should improve as more measurements
are gathered. The impact parameter of Kepler-9\,d is very poorly
constrained by current data, but should also become better determined
in the future. A mass determination for this object has not been made,
and will be challenging given the small amplitude expected for the
reflex motion of the star. The radial-velocity semi-amplitude would be
only about 2.3\,m~s$^{-1}$ assuming a similar mean density as the
Earth. Velocity measurements are further complicated by the presence
of the other two planets in this system.

\begin{deluxetable}{lc}
\tabletypesize{\scriptsize}                                                                                                                           
\tablewidth{0pc} 
\tablecaption{Derived properties of Kepler-9\,d.\label{tab:kepler9d}} 
\tablehead{
\colhead{~~~~~~~~~~~~~~~~~~Parameter~~~~~~~~~~~~~~~~~~} & \colhead{Value\tablenotemark{a}} 
}
\startdata 
Orbital period (days)\dotfill                     & $1.592851\pm0.000045$ \\ [+0.5ex]
Mid-transit epoch (BJD)\dotfill                   & $2,\!455,\!015.0943^{+0.0018}_{-0.0033}$ \\  [+0.5ex]
Orbital semimajor axis (AU)\dotfill               & $0.02730^{+0.00042}_{-0.00043}$ \\ [+0.5ex]
Transit duration (hours)\tablenotemark{b}\dotfill & $1.97^{+0.13}_{-0.17}$ \\  [+0.5ex]
$R_p/R_{\star}$\dotfill                           & $0.0147^{+0.0015}_{-0.0011}$ \\  [+0.5ex]
$R_p$ ($R_{\earth}$)\dotfill                      & $1.64^{+0.19}_{-0.14}$ \\ [+0.5ex] 
$a/R_{\star}$\tablenotemark{c}\dotfill            & $5.54^{+0.51}_{-2.36}$ \\ [+0.5ex]
Equilibrium temperature (K)\tablenotemark{d}\dotfill & $2026 \pm 60$ \\ [-1.5ex]
\enddata
\tablenotetext{a}{Values and uncertainties correspond to the mode and
1-$\sigma$ confidence levels derived from the mode of the \emph{a posteriori}
distributions generated with the Markov Chain Monte Carlo algorithm.}
\tablenotetext{b}{Defined here as the time interval between the first
and last contacts.}  
\tablenotetext{c}{The calculation of the normalized semimajor axis
assumes the orbit is circular.}
\tablenotetext{d}{Zero-albedo equilibrium temperature ignoring the
energy redistribution factor.}

\end{deluxetable}

Many of the most interesting candidates found by \kepler\ will
correspond to Earth-size planets, some of which are expected be in the
habitable zone of their host star. For solar-type stars this implies
reflex motions with radial-velocity amplitudes below current detection
limits, making the spectroscopic measurement of the mass impossible.
Other reasons may also hinder this type of validation, such as rapid
rotation, chromospheric activity, or even the faintness of the star.

The information contained in the \kepler\ light curves on the shape of
a transit-like event is a valuable asset for constraining the vast
range of possible astrophysical false positives that might be
masquerading as a planet. Here we have shown how modeling the light
curve of a candidate directly as a false positive allows to rule out a
significant fraction of blends involving background or foreground
eclipsing binaries or star-planet pairs, as well as hierarchical
triple systems, each of these possibly attenuated by the light of
additional stars. The combination of \blender\ with follow-up
observations consisting of high-resolution imaging, spectroscopy, and
an analysis of the centroid motion of the target leaves no room for a
false positive in the case of Kepler-9\,b and c.  These signals were
previously known to correspond to bona-fide Saturn-size planets
because they display correlated TTVs \citep{Holman:10}. Nevertheless,
the exercise serves to show that, lacking that evidence, it would
still be possible to validate them using the same techniques, thus
supporting the general approach and justifying the application to the
more interesting Kepler-9\,d signal.

Among the advantages of \blender\ for \kepler\ is the ability to
predict the brightness and overall color of a blend in many different
passbands. Brightness information for virtually every \kepler\ target
is available from the KIC in the Sloan $griz$ and 2MASS $JHK_s$ bands,
as well as in the custom D51 passband (518.7\,nm). We have shown
earlier how these can be used to rule out certain blend scenarios that
might otherwise be viable.  The detailed fitting of the photometry
with a false-positive model provides additional discriminating power.
Rough estimates of the properties of a background eclipsing binary
that can mimic a blend have sometimes been made in previous transit
surveys based simply on the apparent brightness of the object and a
representative depth for its undiluted eclipses (such as 50\%). While
such configurations may well reproduce the observed amplitude of a
candidate light curve, not much can be said about the expected shape,
which may be completely wrong. An example of this is seen for
Kepler-9\,d in Figure~\ref{fig:2P}, in which a back-of-the-envelope
calculation of the depth one might predict from the brightness and
$\sim$50\% deep eclipses of this binary may not be far off, but the
detailed shape is not a good match to the observations, and \blender\
easily rules out this scenario giving a poor $\chi^2$ for the
fit. Without these additional constraints on blend properties provided
by the detailed light-curve fitting, the space of parameters open to
false positives would be significantly larger, and more difficult to
exclude by other means.

\acknowledgements

Funding for this Discovery mission is provided by NASA's Science
Mission Directorate.  We are grateful to Leo Girardi for computing
isochrones for this work in the \kepler\ passband, to Frederic Pont
for very helpful discussions on false alarm probabilities, and to
David Sing for advice on limb-darkening coefficients. We also thank
the anonymous referee for insightful comments on the original version
of this paper.

{\it Facilities:} \facility{\kepler\ Mission, Keck\,I (HIRES), WIYN,
Palomar (PHARO)}.


\clearpage


























\begin{landscape}
\begin{deluxetable*}{cccccccccccc}
\tabletypesize{\scriptsize}
\tablewidth{0pc}
\tablecaption{Blend frequency estimate for KOI-377.03 based on stellar densities
and frequencies of eclipsing binaries and transiting
planets.\label{tab:statistics}}
\tablehead{
 & & \multicolumn{4}{c}{Blends involving stellar tertiaries} & & \multicolumn{5}{c}{Blends involving planetary tertiaries} \\ [+0.5ex]
\cline{3-6} \cline{8-12} \\ [-1.5ex]
\colhead{$Kp$ range} &
\colhead{$\Delta Kp$} &
\colhead{Stellar density} &
\colhead{$\rho_{\rm max}$} &
\colhead{Stars} &
\colhead{EBs} & &
\colhead{Stellar density} &
\colhead{$\rho_{\rm max}$} &
\colhead{Stars} &
\colhead{Transiting Jupiters} &
\colhead{Transiting Neptunes}
\\
\colhead{(mag)} &
\colhead{(mag)} &
\colhead{per sq.\ deg} &
\colhead{(\arcsec)} &
\colhead{($\times 10^6$)} &
\colhead{$f_{\rm EB} =0.53$\%} & &
\colhead{per sq.\ deg} &
\colhead{(\arcsec)} &
\colhead{($\times 10^6$)} &
\colhead{6--15\,$R_{\earth}$, $f_{\rm Jup}=0.11$\%} &
\colhead{3.4--6\,$R_{\earth}$, $f_{\rm Nep}=0.10$\%}
\\
\colhead{} &
\colhead{} &
\colhead{} &
\colhead{} &
\colhead{} &
\colhead{($\times 10^6$)} & &
\colhead{} &
\colhead{} &
\colhead{} &
\colhead{($\times 10^6$)} &
\colhead{($\times 10^6$)} \\
\colhead{(1)} &
\colhead{(2)} &
\colhead{(3)} &
\colhead{(4)} &
\colhead{(5)} &
\colhead{(6)} & &
\colhead{(7)} &
\colhead{(8)} &
\colhead{(9)} &
\colhead{(10)} &
\colhead{(11)}
}
\startdata
13.8--14.3  &  0.5 &\nodata &\nodata  & \nodata & \nodata && \nodata&\nodata & \nodata& \nodata & \nodata  \\
14.3--14.8  &  1.0 &\nodata &\nodata  & \nodata & \nodata && \nodata&\nodata & \nodata& \nodata & \nodata  \\
14.8--15.3  &  1.5 &\nodata &\nodata  & \nodata & \nodata && \nodata&\nodata & \nodata& \nodata & \nodata  \\
15.3--15.8  &  2.0 &\nodata &\nodata  & \nodata & \nodata && \nodata&\nodata & \nodata& \nodata & \nodata  \\
15.8--16.3  &  2.5 &\nodata &\nodata  & \nodata & \nodata &&  563   &  0.14  &  2.675 & 0.0029  & 0.0027   \\
16.3--16.8  &  3.0 &\nodata &\nodata  & \nodata & \nodata &&  595   &  0.17  &  4.168 & 0.0046  & 0.0042   \\
16.8--17.3  &  3.5 &\nodata &\nodata  & \nodata & \nodata &&  588   &  0.20  &  5.701 & 0.0063  & 0.0057   \\
17.3--17.8  &  4.0 &\nodata &\nodata  & \nodata & \nodata &&  430   &  0.23  &  5.514 & 0.0061  & 0.0055   \\
17.8--18.3  &  4.5 &\nodata &\nodata  & \nodata & \nodata &&  292   &  0.25  &  4.424 & 0.0049  & 0.0044   \\
18.3--18.8  &  5.0 &   2    &  0.35   &  0.059  & 0.0003  &&  125   &  0.35  &  3.712 & 0.0041  & 0.0037   \\
18.8--19.3  &  5.5 &  14    &  0.45   &  0.687  & 0.0036  &&   57   &  0.45  &  2.798 & 0.0031  & 0.0028   \\
19.3--19.8  &  6.0 &  14    &  0.55   &  1.027  & 0.0054  &&   24   &  0.55  &  1.760 & 0.0019  & 0.0018   \\
19.8--20.3  &  6.5 &  20    &  0.70   &  2.376  & 0.0126  && \nodata&\nodata & \nodata& \nodata & \nodata  \\
20.3--20.8  &  7.0 &  13    &  0.74   &  1.726  & 0.0091  && \nodata&\nodata & \nodata& \nodata & \nodata  \\
20.8--21.3  &  7.5 &   3    &  0.74   &  0.398  & 0.0021  && \nodata&\nodata & \nodata& \nodata & \nodata  \\
21.3--21.8  &  8.0 &   4    &  0.74   &  0.531  & 0.0028  && \nodata&\nodata & \nodata& \nodata & \nodata  \\
21.8--22.3  &  8.5 &   0    &  0.74   &  0.000  & 0.0000  && \nodata&\nodata & \nodata& \nodata & \nodata  \\
22.3--22.8  &  9.0 &   0    &  0.74   &  0.000  & 0.0000  && \nodata&\nodata & \nodata& \nodata & \nodata  \\
\noalign{\vskip 6pt}
\multicolumn{2}{c}{Totals} & 70 &\nodata &  6.804 & {\bf 0.0359} && 2674  &\nodata & 30.752 & {\bf 0.0339}  & {\bf 0.0308}   \\
\noalign{\vskip 4pt}
\hline
\noalign{\vskip 4pt}
\multicolumn{12}{c}{Blend frequency (BF) = $(0.0359 + 0.0339 + 0.0308)\times 10^{-6}= 1.006 \times 10^{-7}$} \\ [-1.0ex]
\enddata
\end{deluxetable*}
\clearpage
\end{landscape}

\end{document}